\documentclass[aps,twocolumn,pr*,amsfonts,showpacs,superscriptaddress,longbibliography]{revtex4-2}
\DeclareUnicodeCharacter{0393}{$\Gamma$}
\usepackage{verbatim,epsfig,amsmath,amssymb,bm,epsf,graphicx,psfrag,bbold,amsthm,amsfonts}
\usepackage[bottom]{footmisc}
\usepackage{hyperref}
\usepackage{cleveref} %should be placed only after hyperref
\usepackage{enumitem}
\usepackage{framed}
\usepackage{mathrsfs}
\usepackage{esint}
\setlist{nosep}
\usepackage{placeins}
\usepackage[all]{xy}
\usepackage{color}
\usepackage[utf8]{inputenc}
\usepackage{float}
\usepackage{natbib}
\usepackage{tikz-cd}
\usepackage{verbatim}
\usepackage{leftidx}
\usepackage{physics}
\usepackage{ulem}

\usepackage{pifont}% http://ctan.org/pkg/pifont

\newcommand{\Db}{{\bm D}}

\newcommand{\thetaq}{\theta_{\bm q}}
\newcommand{\rb}{{\bm r}}
\newcommand{\ub}{{\bm u}}
\newcommand{\kb}{{\bm k}}
\newcommand{\qb}{{\bm q}}
\newcommand{\db}{{\bm d}}
\newcommand{\ab}{{\bm a}}

\newcommand{\gb}{{\bm g}}

\newcommand{\vbold}{{\bm v}}

\newcommand{\Ab}{{\bm A}}

\newcommand{\calA}{\mathcal  A}
\newcommand{\Cbar}{\bar{C}}
\newcommand{\gammacx}{\gamma_{\rm cx}}
\newcommand{\gammasp}{\gamma_{\rm sp}}

\begin{document}
\title{Spontaneous vortex-antivortex lattice and Majorana fermions in \\ rhombohedral graphene}

\author{Filippo Gaggioli}
\email[Email: ]{gfilippo@mit.edu}
\affiliation{Department of Physics, Massachusetts Institute of Technology, Cambridge, MA-02139,USA}

\author{Daniele Guerci}
\email[Email: ]{dguerci@mit.edu}
\affiliation{Department of Physics, Massachusetts Institute of Technology, Cambridge, MA-02139,USA}

\author{Liang Fu}
\email[Email: ]{liangfu@mit.edu}
\affiliation{Department of Physics, Massachusetts Institute of Technology, Cambridge, MA-02139,USA}

\begin{abstract}
The discovery of superconducting states in multilayer rhombohedral graphene with spin and valley polarization \cite{Han_2024} has raised an interesting question: how does superconductivity  cope with time-reversal symmetry breaking?  
In this work, using Ginzburg-Landau theory and microscopic calculation, we predict the existence of a new superconducting state at low electron density, which exhibits a spontaneously formed lattice of vortices and antivortices hosting Majorana zero-modes in their cores. We further identify this vortex-antivortex lattice (VAL) state in the experimental phase diagram and describe its experimental manifestations. 
\end{abstract}
\maketitle

\textit{Introduction --}
It is well-known that time-reversal symmetry plays a fundamental role in conventional $s$-wave superconductivity \cite{Anderson_1959}. It enables electron pairing under weak attraction, underpins the robustness of superconductors against non-magnetic disorder, and explains their fragility against time-reversal symmetry breaking perturbations. For this reason, unconventional superconductivity with broken time-reversal symmetry is  %prime candidates for hosting unconventional—and potentially topological—order parameters. Such systems are of 
highly interesting and long sought after \cite{Kallin_2016,  Maeno_2024}. Moreover, time-reversal-breaking superconductors with chiral order parameters can support Majorana fermions \cite{Volovik_1999, Read_2000, Ivanov_2001, Kozii_2016}, which may be harnessed for topological quantum computing \cite{Nayak_2008_review}.%~\DG{[I feel we should be more specific: not only time-reversal symmetry but also inversion symmetry is broken here. Superconductivity with spin-polarized inversion symmetric electrons is a weak-coupling instability.]} 

It is therefore no surprise that the experimental discovery of superconductivity coexisting with orbital ferromagnetism in multilayer rhombohedral graphene \cite{Han_2024} has sparked widespread excitement across the condensed matter physics community \cite{Chou_2024, Geier_2024, Yoon_2025,Yang_2024,Qin_2024,Jahin_2024,Shavit_2024,Maymann_2025,Parra_2025, JC_ashvin}. In this system, superconductivity is observed at a low electron density (as small as $4\times 10^{11}\,$cm$^{-2}$)  under a large displacement field, which flattens the conduction band edges near $K$ and $K'$ valley. The large density of states give rise to spontaneous spin and valley ferromagnetism which onset at a relatively high temperature. At low temperature, the primary superconducting state (``SC-1'') arises within the fully spin and valley polarized quarter metal, as evidenced by (1) the survival of superconductivity up to very large in-plane magnetic field, (2) a large anomalous Hall effect above $T_c\,\approx 300\, $mK, and (3) quantum oscillations characteristic of a single isospin flavor. 

%The reduction in kinetic energy also amplifies the many-body interactions, enabling the emergence of chiral superconductivity. Moreover, when a single Fermi surface coexists with spin and valley polarization — that is, at moderate displacement fields — rhombohedral graphene becomes a promising candidate for chiral \emph{topological} superconductivity and a viable platform for exploring Majorana zero-modes.

Superconductivity of spin-polarized electrons residing in a single valley is unconventional by all means. Due to Pauli exclusion principle, pairing of single-flavor electrons can only occur with odd orbital angular momentum, such as $p$-wave. %Second, intra-valley Cooper pairing necessarily carries a large crystal momentum $2K$ (although it is irrelevant to the low-energy physics we study below). Last but not the least, 
Furthermore, the absence of time-reversal symmetry in the normal state invites the exploration of new types of superconducting states and phenomena.     

In this work, we predict the existence of a new superconducting phase in rhombohedral graphene at low electron density, which has a spatially modulated pairing order parameter leading to a spontaneously formed lattice of vortices and anti-vortices at zero magnetic field.  As we show by Ginzburg-Landau (GL) theory and microscopic calculation, this state is characterized by a superposition of finite-momentum pairings at three incommensurate wavevectors with the period on the order of $100$nm.

%

% \begin{figure}
%     \includegraphics{Fig_1_edited.pdf}
%     \caption{\textbf{Finite-momentum pairing:} $(a)$ Fermi surface in the $K$ valley, at fixed displacement field and growing electron densities $n_e\approx 0.2, 0.4,0.5$ and $0.6\times 10^{12}\,$cm$^{-2}$.
%     $(b)$ Trigonal warping favors finite momentum pairing that maximizes nesting between the electron (solid line) and hole (dashed) Fermi surfaces.
%     %Lower panel: modulus $|\psi|$ along the vertical black line cut in the main panel.
%     }
%     \label{fig:Fermi_surface}
% \end{figure}

The vortex-antivortex lattice (VAL) state is energetically favored by trigonal warping~\cite{Koshino_2009,Zhang2010} in the band dispersion and 
competes with the $q=0$ uniform state at higher electron density \cite{Geier_2024}. Interestingly, we find that phase transition between these two states can be induced by tuning electron density and out-of-plane magnetic field. Comparing our theoretical phase diagram with experimental data, we identify a region of the likely VAL state in tetralayer graphene, and explain the unusual resistivity behavior in terms of the flux flow associated with ``anomalous'' vortices and anti-vortices.          

%Topological superconductors with odd-pairing host Majorana quasi-particles at their edges. 
% odd-parity pairing gives rise to 
By solving the quasiparticle energy spectrum, we find that spontaneous vortices and anti-vortices host Majorana zero modes, as a result of odd-parity pairing with a single flavor Fermi surface. These Majoranas form two nearly-flat Chern bands near the Fermi level, thus providing a large entropic contribution that further stabilizes the VAL state at finite temperatures.
\begin{figure}
    \includegraphics[width=\linewidth]{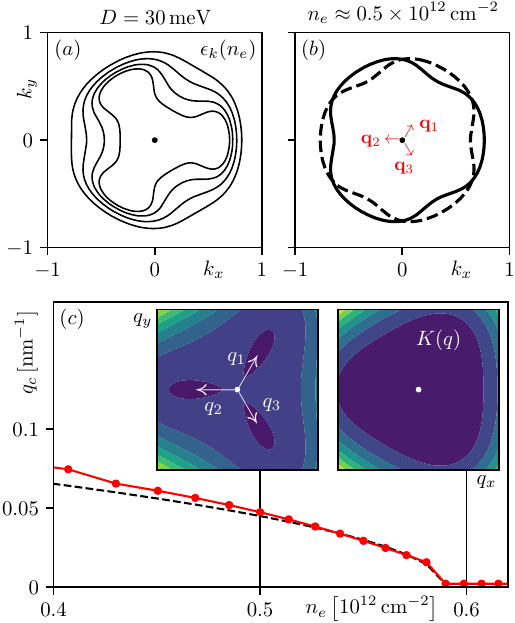}
    \caption{\textbf{Finite-momentum pairing:} Panel $(a)$ shows the Fermi surface in the $K$ valley, at fixed displacement field and growing electron densities $n_e\approx 0.2, 0.4,0.5$ and $0.6\times 10^{12}\,$cm$^{-2}$. Panel $(b)$ displays the electron (solid) and hole (dashed) Fermi surfaces as well as the center-of-mass momentum shifts $\qb_{1,2,3}$. Panel $(c)$ shows the microscopic value of $q_0$ (red) as a function of $n_e$, which displays the approximate scaling $\propto \sqrt{n_{e,0} - n_e}$ (dashed). 
    Inset: colormap of $K(\qb)$ (most negative in purple regions) for $n_e = 0.5$ and $n_e = 0.6\times 10^{12}\,$cm$^{-2}$. }
    \label{fig:Fermi_surface}
\end{figure}
\textit{Spontaneous vortex-antivortex lattice---} We start by developing a Ginzburg-Landau (GL) theory of superconductivity in spin- and valley-polarized rhombohedral graphene based on general symmetry considerations. 
Full isospin polarization restricts the superconducting order parameter to have odd angular momentum $L=\pm1$ ($p$-wave) or $\pm 3$ ($f$-wave). Moreover, 
the valley polarization breaks time-reversal symmetry at the orbital level, even in the absence of an external magnetic field. This lifts the degeneracy between $+L$ and $-L$ pairings, leading to a chiral superconducting state described by  a single scalar order parameter \cite{Geier_2024,Chou_2024}. 
Finally, the valley-polarized normal state preserves three-fold rotation symmetry $C_{3z}$ of rhombohedral graphene and is invariant under the combined operation of reflection $x\rightarrow -x$ and time-reversal, $M_x T$.  
These symmetries are manifested in trigonal warping in the electronic band within a valley. 

Taking into account the above symmetry constraints, the Ginzburg-Landau superconducting free energy takes the form: 
\begin{eqnarray}\label{eq:GL_free_energy}
    F_s = \int &\mathrm{d}^2\rb \; &  \alpha(T) |\psi|^2 + \frac{\hbar^2}{2m^*}|\Db\psi|^2 + \eta|\Db^2\psi|^2 
    \nonumber \\
      & +& \gamma \psi^* \left(D_x^3- 3D_xD^2_y\right)\psi +\cdots,    
\end{eqnarray}
where $\alpha(T) = \alpha(0)(1 - T/T_0)$ with $\alpha(0) < 0$ and $\Db$ is the covariant derivative $\Db=-i\nabla -(2e /\hbar)\Ab$, with $\Ab$ the vector potential. %and $\phi_0 = h/2e$. % the superconducting flux quantum. 

Importantly, the cubic gradient term $\propto \gamma$ (which we refer to as the cubic invariant) is allowed by (1) the presence of the discrete symmetries $C_{3z}$ and $M_xT$ compatible with trigonal warping, and (2) the absence of time-reversal symmetry in the valley polarized state above $T_c$ \cite{Yuan_2022}.
The quartic gradient term with $\eta>0$ is included to prevent the divergence in the gradient energy at large $\gamma$. 
For $\Ab=0$, it is convenient to express the quadratic coefficient $\propto |\psi|^2$ in the free energy in momentum space, where it takes the simple form
\begin{align}\label{eq:propagator}
    K(\bm q) %&= \alpha + cq^2  -d\left(q_x^3 - 3q_xq_y^2\right) + gq^4\nonumber\\
     = \alpha(T) + \hbar^2q^2/2m  -\gamma\,q^3\cos(3\thetaq) + \eta q^4,
\end{align}
with the angle $\thetaq$ specifying the direction of $\qb$. % ($\thetaq=0$ corresponds to $2\bm K$). 

For a sufficiently large value of $\gamma>0$, the gradient term \eqref{eq:propagator} has one minimum at $q = 0$ and three degenerate minima at $\qb_j$ along the directions $\theta_j = (2j-1)\pi/3$ with equal magnitude $|\qb_1|=|\qb_2|=|\qb_3| \equiv q_0$. %$\qb_j = -q_0\mathcal R^{j-1}_{2\pi/3}(1,0)$ with $\mathcal R_{2\pi/3}$ rotation of $120^\circ$ around $z$. 
As $\gamma$ further increases, the global minimum of the free energy \eqref{eq:GL_free_energy} shifts from $q = 0$ to $q_0$, leading to finite momentum superconductivity.
%when $\Delta \alpha = \alpha - \alpha(T)$ is negative. 
This transition from $q=0$ to $q_0$ states 
is first order.  
Our GL theory thus provides a unified description of zero- and finite-momentum states: the latter arises from an instability in gradient energy leading to a spatial modulation of the order parameter.   
%This conclusion drawn from GL theory only relies on the   
%By recasting Eq.\ \eqref{eq:propagator} in a dimensionless form, it can be shown \cite{supplementary} that $q_0$ is proportional to the momentum scale $\sqrt{\hbar^2/2m \eta}$, while $\Delta K_c \propto \eta q_0^4$ -- this will be useful for later calculations.

To determine whether finite momentum superconductivity may actually occur in rhombohedral graphene, % a microscopic foundation for the general symmetry-based result~\eqref{eq:GL_free_energy}, 
we  derive the gradient term $K(\qb)$ from a microscopic calculation using the band dispersion $\epsilon_{\kb}$ for the tetralayer graphene, with $\kb=0$ corresponding to the Dirac point $K$. % and a pairing interaction $V(\rb-\rb')$: %of strength $g$ in a chiral odd-angular-momentum ($L$) channel \cite{Chou_2024,Geier_2024}. 
It reads
\begin{equation}\label{eq:K_q}
    K(\qb) =\frac{1}{2g}-\frac{k_BT}{2}\sum_{\omega_n}\int_{\kb}|\Delta_{\kb}|^2 G_{p_+}G_{p_-},
    % \frac{1}{2g}-\frac{\Omega}{4}\int \frac{d^2\kb}{(2\pi)^2}\frac{|\kb|^2}{\xi_{\qb/2+\kb}+\xi_{\qb/2-\kb}}\\
    %\times\left(\tanh\frac{\xi_{\qb/2+\kb}}{2k_B T}+\tanh\frac{\xi_{\qb/2-\kb}}{2k_B T}\right),
\end{equation}
where $\Delta_\kb$ is the gap function which must be odd in $\kb$: $\Delta_{-\kb}=-\Delta_{\kb}$; $g$ is the corresponding pairing strength; $p_\pm=(\qb/2\pm\kb,\pm \omega_n)$ with $\omega_n=\pi(2n+1)k_BT$, $G_p=1/(i\omega_n-\xi_\kb)$ is the Green's function of free electron, $\xi_{\kb}=\epsilon_{\kb}-\mu$ and $\int_{\kb}=\Omega\int {d^2\kb}/(2\pi)^2$ with $\Omega=\sqrt{3}a^2_G/2$ the unit cell area of graphene. 

As is from Eq.~\eqref{eq:K_q}, the $\qb$ dependence in the gradient term $K(\qb)$ at low temperature is largely determined by the Fermi surface of the valley-polarized normal state,  shown in Fig.~\ref{fig:Fermi_surface}$\,(a)$. 
Two features are noteworthy. 
First, the absence of time reversal and inversion symmetry $\xi_{\kb}\neq \xi_{-\kb}$ removes 
%the Cooper log singularity in $\alpha(T)$ 
the weak-coupling instability, i.e., the log singularity in $\alpha(T)$,
and necessitates a finite interaction strength $g$~\eqref{eq:K_q} for the 
%superconducting instability
superconductivity
to occur, as detailed in the SM~\cite{supplementary}. 
Second, the trigonal warping on the Fermi surface becomes pronounced at low density, resulting in a larger $\gamma$ and triggers a stronger superconducting instability for finite momenta pairing at $\qb_{j}$ due to better particle-particle nesting condition than $q=0$ pairing, as illustrated in Fig.~\ref{fig:Fermi_surface}$\,(b)$. 

The functional form of $K(\qb)$ obtained from our microscopic calculation (assuming a chiral $p$-wave order parameter $\Delta_\kb \propto k_x + ik_y$, as proposed for rhombohedral graphene \cite{Geier_2024}) agrees well with the GL theory~\eqref{eq:propagator}. The absolute value of the pair center-of-mass momentum $q_0$ becomes nonzero below a critical density $n_{e,0}$ and increases %as $\propto\sqrt{n_{e,0} - n_e}$ 
upon further reducing the electron density, as shown in Fig.\ \ref{fig:Fermi_surface}$\,(c)$. Therefore, we conclude that the superconducting state  undergoes a first-order phase transition from being uniform to spatially modulated as the density decreases.

In the presence of pairing instability at three degenerate momenta $\qb_j$ related by symmetry, 
the superconducting state below $T_c$ may be 
a single-$\qb$ state or a multiple-$\qb$ state, depending on the quartic terms in the free energy. 
Restricting to the three dominant $\qb_j$ pairing components $\psi_j$, the free energy density takes the form
\begin{equation}\label{eq:GL_free_energy_density}
    f_s = \alpha \sum_{j=1}^{3}|\psi_j|^2+\beta\sum_{j=1}^{3}|\psi_j|^4+\beta' \sum_{i\neq j}|\psi_i|^2|\psi_j|^2.
\end{equation}
The microscopic calculation of the quartic coefficients gives: 
\begin{equation}\label{quartic_coeff}
\begin{split}
    &\beta=\frac{k_BT}{4}\sum_{\omega_n}\int_{\kb}|\Delta_{\kb}|^4G_{p_+}^2G_{p_-}^2,\\
    &\beta'=\frac{k_BT}{2}\sum_{\omega_n}\int_{\kb}|\Delta_{\kb+\qb_j/2}|^2 |\Delta_{\kb+\qb_l/2}|^2G^2_{p_1}G_{p_2}G_{p_3},\\
\end{split}
\end{equation}
where $p_\pm$ have been defined below Eq.~\eqref{eq:K_q}, and, in the integral for $\beta'$, $p_1=(\kb+(\qb_j+\qb_l)/2,\omega_n)$, $p_{2,3}=(\pm(\qb_l-\qb_j)/2-\kb,-\omega_n)$. 
Performing the integrals in Eq.~\eqref{quartic_coeff} for $\Delta_\kb \propto k_x + ik_y$, we found that $\beta>\beta'>0$, as shown in the SM~\cite{supplementary}, indicating that the triple-$\qb$ state characterized by an equal superposition of $\qb_{1,2,3}$ components is favored over a wide a range of parameters. Its order parameter is 
\begin{align}\label{eq:order_parameter_finite_q}
    \psi =|\psi_c|\left(e^{i\qb_1\cdot\rb} + e^{i\qb_2\cdot\rb}
+e^{i\qb_3\cdot\rb}\right),
\end{align}
with $|\psi_c| = \sqrt{|\alpha|/(2\beta+4\beta')}$ found by minimizing Eq.\ \eqref{eq:GL_free_energy_density}.
% As discussed in the supplementary material \cite{supmat}, the three plane-waves in Eq.\ \eqref{eq:order_parameter_finite_q} may differ for an arbitrary factor $e^{i\phi_i}$. These  phases however play no physical role, and can be gauged away by a redefinition of global phase of the order parameter $\psi$ (1 degree of freedom) and a translation of the reference frame (2 degrees of freedom). 
Note that the relative phases of the three components~\eqref{eq:order_parameter_finite_q} are removed by a gauge transformation involving the global phase of the order parameter and a translation of coordinates~\cite{supplementary}.
%These degrees of freedom are associated to the phase of the order parameter $\chi$ and the phonons $(\phi_1,\phi_2)$.  
%This contrasts with the case of a charge density wave~\cite{Dong_2024,Senthil_2024,Zhang_2024}. %, where one phase continuously tunes the structure from triangular to honeycomb lattice~\cite{Zeng_2024,Zeng_2024sliding}.

\begin{figure}
    \includegraphics{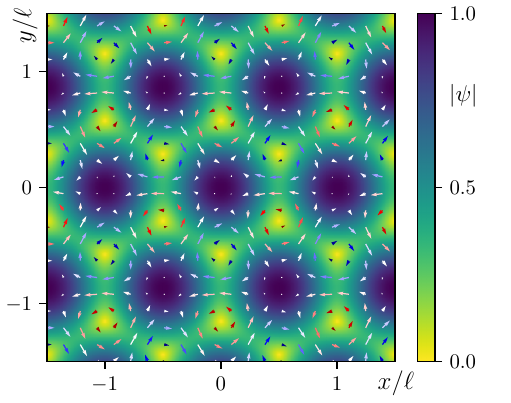}
    \caption{\textbf{Vortex-antivortex lattice}: Order parameter modulus (colormap) and current/phase pattern (arrows) associated to spontaneous VAL state \eqref{eq:order_parameter_finite_q}.
    The coloring of the current arrows reflects the local winding of the order parameter, red for vortex and blue for anti-vortex.
    %Lower panel: modulus $|\psi|$ along the vertical black line cut in the main panel.
    }
    \label{fig1}
\end{figure}

The spatial profile of the modulus and the phase of the triple-$\qb$ order parameter are shown in Fig.~\ref{fig1}. Due to the interference between the three plane-wave components, the order parameter exhibits periodic modulation with a large  period $\ell=4\pi/(3q_0)$ and a honeycomb lattice structure. Furthermore, %$\psi \approx 2\pi i|\psi_c| (x\pm iy)/\ell$ 
$\psi$ has two sets of nodes with $\pm 2\pi$ phase winding around them, which are located at two triangular sublattices $\vbold^{\pm}_{n,m}= n \ab_1 + m \ab_2\pm\ub$,  
where $n,m\in\mathbb Z$, $\ab_{1/2} =\ell\left(\pm 1/2, \sqrt{3}/2\right)$, $\ub = \ell\left(0,1/\sqrt{3}\right)$.
Together, these form a honeycomb lattice of vortices and antivortices, spontaneously formed at zero magnetic field. 
Therefore, we call this spatially modulated superconducting state a vortex-antivortex lattice (VAL) state. 
% The VAL state should be contrasted with other types of finite momentum superconductivity, such as Fulde-Ferrell state with a constant modulus or the Larkin-Ovchinnikov state with a real order parameter.

The finite momentum superconductivity of the VAL state should be contrasted to the Fulde-Ferrell state with a constant modulus or the Larkin-Ovchinnikov state with a real order parameter, and belongs to the more general class of (chiral) pair-density wave (PDWs)  studied in the context of cuprates \cite{Agterberg_2008,Berg_2009,Agterberg_2020_review} and kagome materials \cite{Yin_2022_review, Neupert_2022_review, Deng_2024}.
The topological excitations of PDWs comprise regular vortices, dislocations and combinations of fractional vortices and fractional dislocations \cite{Agterberg_2008,Berg_2009}. Their unbinding via a BKT melting transition can then lead to interesting daughter states with exotic vestigial orders such as charge-$6e$ superconductivity \cite{Agterberg_2011}.

\textit{Experimental manifestation --} 
The VAL state exhibits spontaneous circulating supercurrents and therefore produces a spatially varying magnetic field texture, which can be imaged with SQUID-on-tip scanning techniques \cite{Vasyukov_2013, Uri_2016}.
Using $j\sim e (n_e/d)(\hbar\,q_0/ m)\sim 10^{4}\,$A/cm$^{-2}$ as an estimate \cite{Gaggioli_2024} of the current density circulating in the VAL state ($n_e\sim 10^{12}\,\text{cm}^{-2}$ is the $2$D electron density, $m$ taken to be the bare electron mass and $d\sim 1\,$nm the sample thickness) 
we find magnetic fields $B \sim (4\pi/c)jd$ in the order of $10^{-4}\,$mT, well within the sensitivity range of the experimental technique.

The spatial varying order parameter in VAL phase at zero applied magnetic field displays a single characteristic length scale $\ell$, in contrast with a traditional Abrikosov vortex lattice where the ratio of the vortex core size $\sim\xi$ to the lattice constant $\propto \sqrt{1/H}$ can be tuned by means of the external magnetic field. Nonetheless, in both the VAL and the Abrikosov lattice, an electric current can induce the (thermally assisted) motion of the vortices, creating an electric field inside the superconductor and leading to finite resistivity \cite{Blatter_1994}. 
%This is particularly relevant for rhombohedral graphene, where the interface between different orbital magnetic domains (time-reversed copies where $\gamma, \qb_i,\psi \to -\gamma, -\qb_i, \psi^*$) act as domain walls that can facilitate vortex motion. 
This is particularly relevant for rhombohedral graphene, where disorder and domain walls between different orbital magnetic domains (time-reversed copies where $\gamma, \qb_i,\psi \to -\gamma, -\qb_i, \psi^*$) can make the VAL order short-ranged, facilitating vortex motion.

\begin{figure}
    \includegraphics{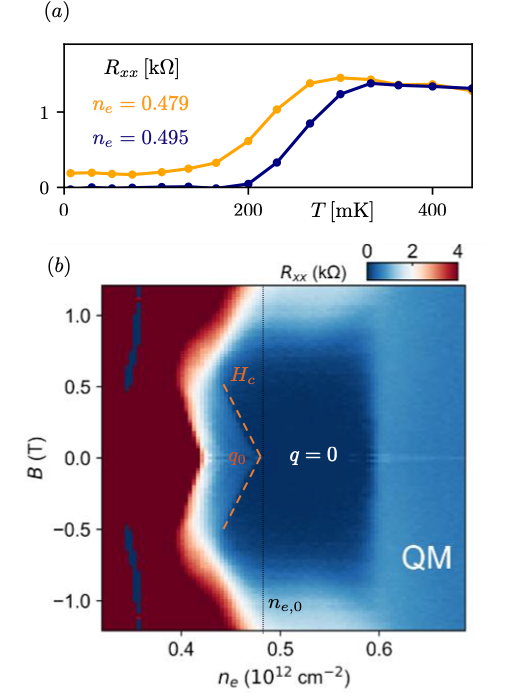}
    \caption{\textbf{Experimental phase diagram from Ref.\ \cite{Han_2024}}:
    $(a)$ Temperature dependence of the longitudinal resistivity in the VAL (orange) and zero-momentum state (blue) when $H=0$. In the VAL state (orange), $R_{xx}$ saturates to a plateau determined by the vortex-antivortex motion.
    $(b)$ At small temperatures and electronic densities $\lesssim 0.48 \times 10^{-12}\,$cm$^{-2}$, the system is in the VAL phase (finite resistivity) for small magnetic fields and jumps into the zero-momentum state (vanishing resistivity) going across the $H_c$ line (orange dashed). For larger fields, the finite-momentum state is metastable and the system prefers the zero-momentum state until the superconductivity vanishes smoothly. 
    }
    \label{fig3}
\end{figure}

In light of our theory, we now examine experimental data on rhombohedral graphene. We note that an anomalous region exists at densities  $n_e\sim 0.4-0.6\times 10^{12}\,$cm$^{-2}$, below SC-1 and above the highly insulating Wigner crystal. Here, the resistivity drops rapidly below a critical temperature (comparable to $T_c$ of SC-1 at slightly higher density) and saturates to a small nonzero value at low temperature, shown in Fig.~\ref{fig3}. This unusual behavior, unaddressed in Ref.\ \cite{Han_2024},  is naturally explained by the flux-flow resistance in our VAL state. %\FG{in the presence of realistic disorder}. 
Drawing on the Bardeen-Stephen formula \cite{Stephen_1965}, we estimate the flux-flow resistivity to be a fraction $(q_0\,\xi)^2\sim 5-10\,\%$ of the normal-state $R_n$, with $\xi \approx 20\,$nm taken from Ref.\ \cite{Han_2024}.  
This explains the low-temperature plateau observed at zero field (panel $(a)$). 
Further evidence for identifying this region as the VAL state is provided by its response to an out-of-plane magnetic field, as will be shown below. 
%Furthermore, the first-order phase transition between the VAL-state and the zero-momentum pairing \FG{quantitatively?} explains the sharp transition between the finite- and zero-resistance states (panel $(a)$) occurring at finite fields for densities $n_e\lesssim \times 10^{12}\,$cm$^{-2}$, as will be shown below.

%Taking  $\xi \approx 20\,$nm from Ref.\ \cite{Han_2024}, we estimate the flux-flow \cite{Blatter_1994} resistance in the VAL-state to be a fraction $(q_0\,\xi)^2\sim 5-10\,\%$ of the normal-state resistance $R_n$. 

% The Maxwell equation $\nabla\times {\bm B} = (4\pi/c) {\bm j}$ with the current density $\sim 10^6\,\text{A/cm}^2$ and thickness $\sim 1\,$nm typical of $2$D superconductors, we estimate that the VAV lattice should produce a magnetic field $\sim 10^{-3}\,$mT well accessible to SQUID-on-tip imaging techniques \cite{Vasyukov_2013, Uri_2016}.

 %

%

\textit{VAL under out-of-plane magnetic field -- }
We now proceed to discuss the response of the VAL state in rhombohedral graphene to an out-of-plane magnetic field.
Conventional $q=0$ superconductivity is destroyed at $H_{c2}=\Phi_0/(2\pi \xi^2)$ when vortices strongly overlap. 
However, for the VAL state, 
%Using $\ell_B^{-1} = \sqrt{\phi_0/2\pi H}$ as the typical momentum scale associated to the magnetic field, we can see that the kernel \eqref{eq:propagator} vanishes and the system returns to the normal state for fields where $\ell_B^{-1}\ll q_0$, provided that the temperature is sufficiently low and $|\Delta K_c| \ll |K_c|$.
a small magnetic field $\sim\phi_0 q_0^2$ is enough to penalize the energetics of the VAL and induce a first order phase transition into the conventional superconducting state, as we will now show.

This transition is marked by a change in the spatial profile of the order parameter driven by the magnetic field. 
To estimate the occurrence of the transition, we can account for the effect of $H$ via a momentum shift of the order of  the inverse magnetic length $\ell_B^{-1}=\sqrt{\phi_0/2\pi H}$. 
Imposing that $K(q_0 + \ell_B^{-1})$ is equal to the barrier $\sim \eta q_0^4$ separating the finite- and zero-momentum states, and using that the curvature of $K(\qb)$ around $\qb_i$ scales with $\eta q_0^2$ (cf SM \cite{supplementary}), we then find that $H_c \sim \phi_0 q_0^2$. Using the dependence of $q_0$ on the electron density shown from Fig.\ \ref{fig:Fermi_surface}$\,(c)$, we find the critical field displays an approximate linear scaling with $n_{e0}-n_e$; this result reproduces the observed phase boundary in Fig.\ \ref{fig3}$\,(b)$.

% then obtain
% It can be seen from the mean field solution to $K(\Db) \psi(\rb)=0$ with magnetic field present. for the finite-momentum state becomes comparable to the barrier $|\Delta K_c|$, under the effect of a magnetic field. Expanding the kernel \eqref{eq:propagator} around the minima at $\qb_i$, we write the above condition as
% %
% \begin{equation}\label{eq:propagator_twist}
%  \Cbar \left(1-a\cos2(\theta_i-\theta_{\delta\qb})\right)\delta q^2 \approx |\Delta K_c| ,
% %\delta K_i(\kb)  \approx C_\parallel \left(\frac{\kb\cdot\qb_i}{q_0}\right)^2 + C_\perp \left(\frac{\kb\wedge\qb_i}{q_0}\right)^2,
% \end{equation}
% %
% where the coefficients $\Cbar > 0$ and $4/5 \geq a > 1/2$ that characterize, respectively, the average curvature and the anisotropy around $K(\qb_i)$. The curvature $\Cbar$ scales with $\eta q_0^2$, indicating that $|\Delta K_c|/\Cbar \sim q_0^2$ \cite{supplementary}.

% Treating \eqref{eq:propagator_twist} as the equation for an elliptical orbit in momentum space, the critical field $H_c$ can be estimated as \cite{Onsager_1952}
% %
% \begin{equation}\label{eq:H_c}
%      H_c\sim  \frac{\phi_0 \,q_0^2}{\sqrt{1-a^2}}.
% \end{equation}
% %
% At $H_c$ the VAL disappears via a first order transition and the resistivity jumps to zero, explaining the steep decrease observed in the experiment upon increasing the magnetic field \cite{Han_2024}.
% Further increasing $H$ beyond $H_c$, the uniform superconductivity is smoothly suppressed until the system returns to normal when $K(\ell_B^{-1}) \sim 0$.

%
\begin{figure}
    \includegraphics{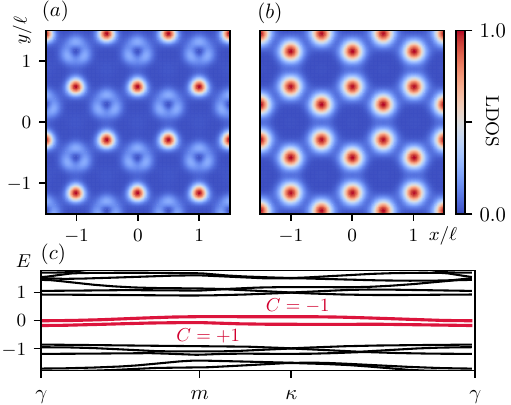}
    \caption{\textbf{Majorana zero-modes}: Local density of states for the  $C=-1$ $(a)$ and for the combined $C=\pm 1$ Majorana bands $(b)$ as measured by a local STM probe. As expected for a topological superconductor, the MZMs are concentrated in the cores of the VAL.
    $(c)$ Energy dispersion of the BdG bands, expressed in units of the gap. The Majorana bands (red lines) are in gap states with Chern number $C=\pm1$ and energy spacing $\sim W$. 
    }
    \label{fig2Majorana}
\end{figure}

\textit{Majoranas and Pomeranchuk effect --} Topological superconductors host Majorana zero-modes (MZMs) inside their vortex cores \cite{Read_2000,Ivanov_2001}. These exotic particles manifest themselves as in-gap states with zero energy and display non-Abelian statistics, thus holding the promise to fault-tolerant topological quantum computing \cite{Nayak_2008_review}. 
Inspired by the presence of a spontaneous VAL inside our system, we derive a microscopic Bogoliubov-de Gennes Hamiltonian $h_\text{BdG}$ for the VAL phase~\cite{Franz_2000,Vafek_2001,Chiu_2015,Liu_2015,Pu_2024,supplementary} which in the Nambu basis $(\Psi,\Psi^\dagger)$ reads:
\begin{equation}
    h_{\rm BdG}(\rb)= \begin{pmatrix}
        \xi_{-i\nabla} & \{\psi(\rb),-i\bar \partial\}/2\\
        h.c. & -\xi_{i\nabla} 
    \end{pmatrix},
\end{equation}
where $\xi_{-i\nabla}=\epsilon_{-i\nabla}-\mu$ and $\bar\partial=\partial_x+i\partial_y$ captures the chiral character of the $p$-wave superconductor. 

The spectrum of the Hamiltonian $h_\text{BdG}$ is influenced by the order parameter $\psi(\rb)$ and, for a sufficiently large magnitude $|\psi_c|$, is gapped everywhere in momentum space (at small $|\psi_c|$, a Bogoliubov-Fermi surface is expected \cite{Brydon_2018,Yuan_2018}).
For a single-$q$ condensate $|\psi(\rb)|$ is constant.  %and the quasiparticle spectrum is generically gapped with the exception of the \FG{Bogoliubov-Fermi surfaces that appear when the superconductivity is weak.}
%nodal configurations where electron-hole surfaces intersect at $\kb=0$.
% 
On the other hand, in the VAL, $|\psi(\rb)|$ is spatially modulated and vanishes as $\psi(\rb+ \bm v_{n,m}^+)\propto z$ and $\psi(\rb+ \bm v_{n,m}^-)\propto z^*$ with $z=x+iy$ around the vortices and antivortices, respectively, forming the honeycomb lattice in Fig.~\ref{fig1}.  

Interestingly, Majorana zero-modes are realized~\cite{Read_2000} in proximity of these nodes. The periodic array characterizing the VAL gives rise to in-gap flat bands of Majorana fermions (red lines in Fig.~\ref{fig2Majorana}$\,(c)$). 
The corresponding local density of state (LDOS) is concentrated inside the cores of the VAL, as shown in Fig.\ \ref{fig2Majorana}$\,(a)$ and $(b)$. The two panels display the LDOS measured as by an STM probe at small positive bias, at zero and finite temperatures. 
In the zero-temperature case, quasi-particles fill all the BdG bands with negative energies and electrons can only tunnel in the superconductor via the particle sector of the positive Majorana band. 
Because of the intrinsic angular momentum of the chiral Cooper pairs, the quasiparticles in the cores of vortices and antivortices have different 
wavefunctions~\cite{Kraus_2009}, resulting in a  different LDOS for the  $\vbold^{\pm}$ sublattices (Fig.\ \ref{fig2Majorana}$\,(a)$).
%The resulting LDOS distinguishes between vortices and antivortices~\cite{Kraus_2009} (Fig.\ \ref{fig2Majorana}$\,(a)$). 
When the temperature is larger than the Majorana bands spacing $\sim W$, on the other hand, occupation of the two Majorana bands is approximately symmetric and the electrons can tunnel as a particles in the upper band and holes in the lower band, resulting in a symmetric LDOS (Fig.~\ref{fig2Majorana}$\,(b)$). 

The two Majorana bands are separated by a topological gap and have opposite Chern number $C=\pm 1$. This gap is opened by the time-reversal symmetry breaking of the next-to-nearest neighbor Majorana hoppings on the honeycomb VAL, similarly to what happens in the Haldane-Kitaev honeycomb model \cite{Haldane_1988,Kitaev_2006}.

Remarkably rich physics emerges upon considering the physics of Majorana zero-modes at finite-temperatures. Because of MZMs, a spontaneous VAL with $2N$ vortex cores has a $2^N$-fold degeneracy \cite{Ivanov_2001}, giving rise to an entropic contribution to the superconducting free energy that scales as $-TN\log 2 \sim -T (\calA/\calA_{\rm uc})$ and is extensive in the sample area $\calA$. At finite temperatures, the MZMs therefore give rise to a \textit{Pomeranchuk effect} \cite{Pomeranchuk_1950} that reduces the free energy of the VAL state, further stabilizing it against other possible order parameter configurations. While this effect was previously observed in magic-angle twisted graphene~\cite{Rozen_2021,Saito_2021}, where it is of magnetic origin, the topological root of the Pomeranchuk effect in rhombohedral graphene makes it unprecedented.

\textit{Discussion --} Spin and valley polarized superconductivity is one of the most fascinating and challenging phases recently realized in multilayer graphene materials~\cite{Han_2024}. 
Our work presents a new state of matter, the spontaneous vortex-antivortex honeycomb lattice, that is realized in the low-electron-density regime of rhombohedral graphene.
In a range of displacement fields, the cores of the VAL host Majorana fermions that form in-gap flat bands with topological properties.
%In the low-electron-density regime of rhombohedral graphene, our microscopic calculations demonstrate that trigonal warping favors a finite center-of-mass momentum, while the quartic term selects an equal superposition of the three condensates related by threefold rotational symmetry. 
%This superposition features nodes on an honeycomb lattice with sublattices featuring opposite vorticity and, therefore, realizes in-gap flat bands of Majorana fermions.  
%

The VAL shows a number of phenomenological properties in good agreement with the recent observations~\cite{Han_2024}. First, a current flowing in the sample induces motion of vortices and antivortices resulting in a finite residual resistance. Second, an out-of-plane magnetic field induces a first-order transition from the VAL to a uniform superconductor, consistently with the observation of a zero-resistance state stabilized at finite field. 
Finally, our theory predicts a Pomeranchuk effect that arises from the degeneracy of the Majorana zero-modes, which further stabilizes the VAL phase over the uniform superconductor as temperature increases. 

We propose real-space imaging of the VAL via STM to detect the Majorana fermions, and experiments with microwave frequencies to investigate the optical phonon mode~\cite{Zhang_1993,Gabay_1993} naturally associated to the VAL bipartite structure. 
Finally, we hope that our work will motivate experimental efforts to directly image the magnetic texture associated to the spontaneous vortices and antivortices, for example with the SQUID-on-tip scanning technique.

\textit{ Acknowledgments --}
It is our pleasure to thank Long Ju and Tonghang Han for insightful discussions and providing us the experimental figures used in this work. We also thank Yoichi Ando for a stimulating remark about Ref.\ \cite{Han_2024} and  Max Geier for discussion in the early stage of this project.  
This work was supported by a Simons Investigator Award from the Simons Foundation. F.G. is grateful for the financial support from the Swiss National Science Foundation (Postdoc.Mobility Grant No. 222230). 

\textit{Note Added --} Two recent preprints also discuss finite-momentum superconductivity in the context of rhombohedral graphene~\cite{Christos_2025, Gil_2025}.

\nocite{Read_2000,Franz_2000,Vafek_2001,Jung2013,Zhang2010,Dong_2024,Geier_2024}
\bibliography{biblio.bib}

\begin{thebibliography}{54}%
\makeatletter
\providecommand \@ifxundefined [1]{%
 \@ifx{#1\undefined}
}%
\providecommand \@ifnum [1]{%
 \ifnum #1\expandafter \@firstoftwo
 \else \expandafter \@secondoftwo
 \fi
}%
\providecommand \@ifx [1]{%
 \ifx #1\expandafter \@firstoftwo
 \else \expandafter \@secondoftwo
 \fi
}%
\providecommand \natexlab [1]{#1}%
\providecommand \enquote  [1]{``#1''}%
\providecommand \bibnamefont  [1]{#1}%
\providecommand \bibfnamefont [1]{#1}%
\providecommand \citenamefont [1]{#1}%
\providecommand \href@noop [0]{\@secondoftwo}%
\providecommand \href [0]{\begingroup \@sanitize@url \@href}%
\providecommand \@href[1]{\@@startlink{#1}\@@href}%
\providecommand \@@href[1]{\endgroup#1\@@endlink}%
\providecommand \@sanitize@url [0]{\catcode `\\12\catcode `\$12\catcode `\&12\catcode `\#12\catcode `\^12\catcode `\_12\catcode `\%12\relax}%
\providecommand \@@startlink[1]{}%
\providecommand \@@endlink[0]{}%
\providecommand \url  [0]{\begingroup\@sanitize@url \@url }%
\providecommand \@url [1]{\endgroup\@href {#1}{\urlprefix }}%
\providecommand \urlprefix  [0]{URL }%
\providecommand \Eprint [0]{\href }%
\providecommand \doibase [0]{https://doi.org/}%
\providecommand \selectlanguage [0]{\@gobble}%
\providecommand \bibinfo  [0]{\@secondoftwo}%
\providecommand \bibfield  [0]{\@secondoftwo}%
\providecommand \translation [1]{[#1]}%
\providecommand \BibitemOpen [0]{}%
\providecommand \bibitemStop [0]{}%
\providecommand \bibitemNoStop [0]{.\EOS\space}%
\providecommand \EOS [0]{\spacefactor3000\relax}%
\providecommand \BibitemShut  [1]{\csname bibitem#1\endcsname}%
\let\auto@bib@innerbib\@empty
%</preamble>
\bibitem [{\citenamefont {Han}\ \emph {et~al.}(2024)\citenamefont {Han}, \citenamefont {Lu}, \citenamefont {Yao}, \citenamefont {Shi}, \citenamefont {Yang}, \citenamefont {Seo}, \citenamefont {Ye}, \citenamefont {Wu}, \citenamefont {Zhou}, \citenamefont {Liu}, \citenamefont {Shi}, \citenamefont {Hua}, \citenamefont {Watanabe}, \citenamefont {Taniguchi}, \citenamefont {Xiong}, \citenamefont {Fu},\ and\ \citenamefont {Ju}}]{Han_2024}%
  \BibitemOpen
  \bibfield  {author} {\bibinfo {author} {\bibfnamefont {T.}~\bibnamefont {Han}}, \bibinfo {author} {\bibfnamefont {Z.}~\bibnamefont {Lu}}, \bibinfo {author} {\bibfnamefont {Y.}~\bibnamefont {Yao}}, \bibinfo {author} {\bibfnamefont {L.}~\bibnamefont {Shi}}, \bibinfo {author} {\bibfnamefont {J.}~\bibnamefont {Yang}}, \bibinfo {author} {\bibfnamefont {J.}~\bibnamefont {Seo}}, \bibinfo {author} {\bibfnamefont {S.}~\bibnamefont {Ye}}, \bibinfo {author} {\bibfnamefont {Z.}~\bibnamefont {Wu}}, \bibinfo {author} {\bibfnamefont {M.}~\bibnamefont {Zhou}}, \bibinfo {author} {\bibfnamefont {H.}~\bibnamefont {Liu}}, \bibinfo {author} {\bibfnamefont {G.}~\bibnamefont {Shi}}, \bibinfo {author} {\bibfnamefont {Z.}~\bibnamefont {Hua}}, \bibinfo {author} {\bibfnamefont {K.}~\bibnamefont {Watanabe}}, \bibinfo {author} {\bibfnamefont {T.}~\bibnamefont {Taniguchi}}, \bibinfo {author} {\bibfnamefont {P.}~\bibnamefont {Xiong}}, \bibinfo {author} {\bibfnamefont {L.}~\bibnamefont {Fu}},\ and\ \bibinfo {author} {\bibfnamefont
  {L.}~\bibnamefont {Ju}},\ }\href {https://arxiv.org/abs/2408.15233} {\bibinfo {title} {Signatures of chiral superconductivity in rhombohedral graphene}} (\bibinfo {year} {2024}),\ \Eprint {https://arxiv.org/abs/2408.15233} {arXiv:2408.15233 [cond-mat.mes-hall]} \BibitemShut {NoStop}%
\bibitem [{\citenamefont {Anderson}(1959)}]{Anderson_1959}%
  \BibitemOpen
  \bibfield  {author} {\bibinfo {author} {\bibfnamefont {P.}~\bibnamefont {Anderson}},\ }\bibfield  {title} {\bibinfo {title} {Theory of dirty superconductors},\ }\href {https://doi.org/https://doi.org/10.1016/0022-3697(59)90036-8} {\bibfield  {journal} {\bibinfo  {journal} {Journal of Physics and Chemistry of Solids}\ }\textbf {\bibinfo {volume} {11}},\ \bibinfo {pages} {26} (\bibinfo {year} {1959})}\BibitemShut {NoStop}%
\bibitem [{Kal(2016)}]{Kallin_2016}%
  \BibitemOpen
  \bibfield  {title} {\bibinfo {title} {Chiral superconductors},\ }\href {https://dx.doi.org/10.1088/0034-4885/79/5/054502} {\bibfield  {journal} {\bibinfo  {journal} {Reports on Progress in Physics}\ }\textbf {\bibinfo {volume} {79}},\ \bibinfo {pages} {054502} (\bibinfo {year} {2016})}\BibitemShut {NoStop}%
\bibitem [{\citenamefont {Maeno}\ \emph {et~al.}(2024)\citenamefont {Maeno}, \citenamefont {Ikeda},\ and\ \citenamefont {Mattoni}}]{Maeno_2024}%
  \BibitemOpen
  \bibfield  {author} {\bibinfo {author} {\bibfnamefont {Y.}~\bibnamefont {Maeno}}, \bibinfo {author} {\bibfnamefont {A.}~\bibnamefont {Ikeda}},\ and\ \bibinfo {author} {\bibfnamefont {G.}~\bibnamefont {Mattoni}},\ }\bibfield  {title} {\bibinfo {title} {Thirty years of puzzling superconductivity in sr2ruo4},\ }\href {https://doi.org/10.1038/s41567-024-02656-0} {\bibfield  {journal} {\bibinfo  {journal} {Nature Physics}\ }\textbf {\bibinfo {volume} {20}},\ \bibinfo {pages} {1712} (\bibinfo {year} {2024})}\BibitemShut {NoStop}%
\bibitem [{\citenamefont {Volovik}(1999)}]{Volovik_1999}%
  \BibitemOpen
  \bibfield  {author} {\bibinfo {author} {\bibfnamefont {G.~E.}\ \bibnamefont {Volovik}},\ }\bibfield  {title} {\bibinfo {title} {Fermion zero modes on vortices in chiral superconductors},\ }\href {https://doi.org/10.1134/1.568223} {\bibfield  {journal} {\bibinfo  {journal} {Journal of Experimental and Theoretical Physics Letters}\ }\textbf {\bibinfo {volume} {70}},\ \bibinfo {pages} {609} (\bibinfo {year} {1999})}\BibitemShut {NoStop}%
\bibitem [{\citenamefont {Read}\ and\ \citenamefont {Green}(2000)}]{Read_2000}%
  \BibitemOpen
  \bibfield  {author} {\bibinfo {author} {\bibfnamefont {N.}~\bibnamefont {Read}}\ and\ \bibinfo {author} {\bibfnamefont {D.}~\bibnamefont {Green}},\ }\bibfield  {title} {\bibinfo {title} {Paired states of fermions in two dimensions with breaking of parity and time-reversal symmetries and the fractional quantum hall effect},\ }\href {https://doi.org/10.1103/PhysRevB.61.10267} {\bibfield  {journal} {\bibinfo  {journal} {Phys. Rev. B}\ }\textbf {\bibinfo {volume} {61}},\ \bibinfo {pages} {10267} (\bibinfo {year} {2000})}\BibitemShut {NoStop}%
\bibitem [{\citenamefont {Ivanov}(2001)}]{Ivanov_2001}%
  \BibitemOpen
  \bibfield  {author} {\bibinfo {author} {\bibfnamefont {D.~A.}\ \bibnamefont {Ivanov}},\ }\bibfield  {title} {\bibinfo {title} {Non-abelian statistics of half-quantum vortices in $\mathit{p}$-wave superconductors},\ }\href {https://doi.org/10.1103/PhysRevLett.86.268} {\bibfield  {journal} {\bibinfo  {journal} {Phys. Rev. Lett.}\ }\textbf {\bibinfo {volume} {86}},\ \bibinfo {pages} {268} (\bibinfo {year} {2001})}\BibitemShut {NoStop}%
\bibitem [{\citenamefont {Kozii}\ \emph {et~al.}(2016)\citenamefont {Kozii}, \citenamefont {Venderbos},\ and\ \citenamefont {Fu}}]{Kozii_2016}%
  \BibitemOpen
  \bibfield  {author} {\bibinfo {author} {\bibfnamefont {V.}~\bibnamefont {Kozii}}, \bibinfo {author} {\bibfnamefont {J.~W.~F.}\ \bibnamefont {Venderbos}},\ and\ \bibinfo {author} {\bibfnamefont {L.}~\bibnamefont {Fu}},\ }\bibfield  {title} {\bibinfo {title} {Three-dimensional majorana fermions in chiral superconductors},\ }\href {https://doi.org/10.1126/sciadv.1601835} {\bibfield  {journal} {\bibinfo  {journal} {Science Advances}\ }\textbf {\bibinfo {volume} {2}},\ \bibinfo {pages} {e1601835} (\bibinfo {year} {2016})},\ \Eprint {https://arxiv.org/abs/https://www.science.org/doi/pdf/10.1126/sciadv.1601835} {https://www.science.org/doi/pdf/10.1126/sciadv.1601835} \BibitemShut {NoStop}%
\bibitem [{\citenamefont {Nayak}\ \emph {et~al.}(2008)\citenamefont {Nayak}, \citenamefont {Simon}, \citenamefont {Stern}, \citenamefont {Freedman},\ and\ \citenamefont {Das~Sarma}}]{Nayak_2008_review}%
  \BibitemOpen
  \bibfield  {author} {\bibinfo {author} {\bibfnamefont {C.}~\bibnamefont {Nayak}}, \bibinfo {author} {\bibfnamefont {S.~H.}\ \bibnamefont {Simon}}, \bibinfo {author} {\bibfnamefont {A.}~\bibnamefont {Stern}}, \bibinfo {author} {\bibfnamefont {M.}~\bibnamefont {Freedman}},\ and\ \bibinfo {author} {\bibfnamefont {S.}~\bibnamefont {Das~Sarma}},\ }\bibfield  {title} {\bibinfo {title} {Non-abelian anyons and topological quantum computation},\ }\href {https://doi.org/10.1103/RevModPhys.80.1083} {\bibfield  {journal} {\bibinfo  {journal} {Rev. Mod. Phys.}\ }\textbf {\bibinfo {volume} {80}},\ \bibinfo {pages} {1083} (\bibinfo {year} {2008})}\BibitemShut {NoStop}%
\bibitem [{\citenamefont {Chou}\ \emph {et~al.}(2024)\citenamefont {Chou}, \citenamefont {Zhu},\ and\ \citenamefont {Sarma}}]{Chou_2024}%
  \BibitemOpen
  \bibfield  {author} {\bibinfo {author} {\bibfnamefont {Y.-Z.}\ \bibnamefont {Chou}}, \bibinfo {author} {\bibfnamefont {J.}~\bibnamefont {Zhu}},\ and\ \bibinfo {author} {\bibfnamefont {S.~D.}\ \bibnamefont {Sarma}},\ }\href {https://arxiv.org/abs/2409.06701} {\bibinfo {title} {Intravalley spin-polarized superconductivity in rhombohedral tetralayer graphene}} (\bibinfo {year} {2024}),\ \Eprint {https://arxiv.org/abs/2409.06701} {arXiv:2409.06701 [cond-mat.supr-con]} \BibitemShut {NoStop}%
\bibitem [{\citenamefont {Geier}\ \emph {et~al.}(2024)\citenamefont {Geier}, \citenamefont {Davydova},\ and\ \citenamefont {Fu}}]{Geier_2024}%
  \BibitemOpen
  \bibfield  {author} {\bibinfo {author} {\bibfnamefont {M.}~\bibnamefont {Geier}}, \bibinfo {author} {\bibfnamefont {M.}~\bibnamefont {Davydova}},\ and\ \bibinfo {author} {\bibfnamefont {L.}~\bibnamefont {Fu}},\ }\href {https://arxiv.org/abs/2409.13829} {\bibinfo {title} {Chiral and topological superconductivity in isospin polarized multilayer graphene}} (\bibinfo {year} {2024}),\ \Eprint {https://arxiv.org/abs/2409.13829} {arXiv:2409.13829 [cond-mat.supr-con]} \BibitemShut {NoStop}%
\bibitem [{\citenamefont {Yoon}\ \emph {et~al.}(2025)\citenamefont {Yoon}, \citenamefont {Xu}, \citenamefont {Barlas},\ and\ \citenamefont {Zhang}}]{Yoon_2025}%
  \BibitemOpen
  \bibfield  {author} {\bibinfo {author} {\bibfnamefont {C.}~\bibnamefont {Yoon}}, \bibinfo {author} {\bibfnamefont {T.}~\bibnamefont {Xu}}, \bibinfo {author} {\bibfnamefont {Y.}~\bibnamefont {Barlas}},\ and\ \bibinfo {author} {\bibfnamefont {F.}~\bibnamefont {Zhang}},\ }\href {https://arxiv.org/abs/2502.17555} {\bibinfo {title} {Quarter metal superconductivity}} (\bibinfo {year} {2025}),\ \Eprint {https://arxiv.org/abs/2502.17555} {arXiv:2502.17555 [cond-mat.mes-hall]} \BibitemShut {NoStop}%
\bibitem [{\citenamefont {Yang}\ and\ \citenamefont {Zhang}(2024)}]{Yang_2024}%
  \BibitemOpen
  \bibfield  {author} {\bibinfo {author} {\bibfnamefont {H.}~\bibnamefont {Yang}}\ and\ \bibinfo {author} {\bibfnamefont {Y.-H.}\ \bibnamefont {Zhang}},\ }\href {https://arxiv.org/abs/2411.02503} {\bibinfo {title} {Topological incommensurate fulde-ferrell-larkin-ovchinnikov superconductor and bogoliubov fermi surface in rhombohedral tetra-layer graphene}} (\bibinfo {year} {2024}),\ \Eprint {https://arxiv.org/abs/2411.02503} {arXiv:2411.02503 [cond-mat.supr-con]} \BibitemShut {NoStop}%
\bibitem [{\citenamefont {Qin}\ and\ \citenamefont {Wu}(2024)}]{Qin_2024}%
  \BibitemOpen
  \bibfield  {author} {\bibinfo {author} {\bibfnamefont {Q.}~\bibnamefont {Qin}}\ and\ \bibinfo {author} {\bibfnamefont {C.}~\bibnamefont {Wu}},\ }\href {https://arxiv.org/abs/2412.07145} {\bibinfo {title} {Chiral finite-momentum superconductivity in the tetralayer graphene}} (\bibinfo {year} {2024}),\ \Eprint {https://arxiv.org/abs/2412.07145} {arXiv:2412.07145 [cond-mat.supr-con]} \BibitemShut {NoStop}%
\bibitem [{\citenamefont {Jahin}\ and\ \citenamefont {Lin}(2024)}]{Jahin_2024}%
  \BibitemOpen
  \bibfield  {author} {\bibinfo {author} {\bibfnamefont {A.}~\bibnamefont {Jahin}}\ and\ \bibinfo {author} {\bibfnamefont {S.-Z.}\ \bibnamefont {Lin}},\ }\href {https://arxiv.org/abs/2411.09664} {\bibinfo {title} {Enhanced kohn-luttinger topological superconductivity in bands with nontrivial geometry}} (\bibinfo {year} {2024}),\ \Eprint {https://arxiv.org/abs/2411.09664} {arXiv:2411.09664 [cond-mat.supr-con]} \BibitemShut {NoStop}%
\bibitem [{\citenamefont {Shavit}\ and\ \citenamefont {Alicea}(2024)}]{Shavit_2024}%
  \BibitemOpen
  \bibfield  {author} {\bibinfo {author} {\bibfnamefont {G.}~\bibnamefont {Shavit}}\ and\ \bibinfo {author} {\bibfnamefont {J.}~\bibnamefont {Alicea}},\ }\href {https://arxiv.org/abs/2411.05071} {\bibinfo {title} {Quantum geometric unconventional superconductivity}} (\bibinfo {year} {2024}),\ \Eprint {https://arxiv.org/abs/2411.05071} {arXiv:2411.05071 [cond-mat.supr-con]} \BibitemShut {NoStop}%
\bibitem [{\citenamefont {May-Mann}\ \emph {et~al.}(2025)\citenamefont {May-Mann}, \citenamefont {Helbig},\ and\ \citenamefont {Devakul}}]{Maymann_2025}%
  \BibitemOpen
  \bibfield  {author} {\bibinfo {author} {\bibfnamefont {J.}~\bibnamefont {May-Mann}}, \bibinfo {author} {\bibfnamefont {T.}~\bibnamefont {Helbig}},\ and\ \bibinfo {author} {\bibfnamefont {T.}~\bibnamefont {Devakul}},\ }\href {https://arxiv.org/abs/2503.05697} {\bibinfo {title} {How pairing mechanism dictates topology in valley-polarized superconductors with berry curvature}} (\bibinfo {year} {2025}),\ \Eprint {https://arxiv.org/abs/2503.05697} {arXiv:2503.05697 [cond-mat.supr-con]} \BibitemShut {NoStop}%
\bibitem [{\citenamefont {Parra-Martinez}\ \emph {et~al.}(2025)\citenamefont {Parra-Martinez}, \citenamefont {Jimeno-Pozo}, \citenamefont {Phong}, \citenamefont {Sainz-Cruz}, \citenamefont {Kaplan}, \citenamefont {Emanuel}, \citenamefont {Oreg}, \citenamefont {Pantaleon}, \citenamefont {Silva-Guillen},\ and\ \citenamefont {Guinea}}]{Parra_2025}%
  \BibitemOpen
  \bibfield  {author} {\bibinfo {author} {\bibfnamefont {G.}~\bibnamefont {Parra-Martinez}}, \bibinfo {author} {\bibfnamefont {A.}~\bibnamefont {Jimeno-Pozo}}, \bibinfo {author} {\bibfnamefont {V.~T.}\ \bibnamefont {Phong}}, \bibinfo {author} {\bibfnamefont {H.}~\bibnamefont {Sainz-Cruz}}, \bibinfo {author} {\bibfnamefont {D.}~\bibnamefont {Kaplan}}, \bibinfo {author} {\bibfnamefont {P.}~\bibnamefont {Emanuel}}, \bibinfo {author} {\bibfnamefont {Y.}~\bibnamefont {Oreg}}, \bibinfo {author} {\bibfnamefont {P.~A.}\ \bibnamefont {Pantaleon}}, \bibinfo {author} {\bibfnamefont {J.~A.}\ \bibnamefont {Silva-Guillen}},\ and\ \bibinfo {author} {\bibfnamefont {F.}~\bibnamefont {Guinea}},\ }\href {https://arxiv.org/abs/2502.19474} {\bibinfo {title} {Band renormalization, quarter metals, and chiral superconductivity in rhombohedral tetralayer graphene}} (\bibinfo {year} {2025}),\ \Eprint {https://arxiv.org/abs/2502.19474} {arXiv:2502.19474 [cond-mat.str-el]} \BibitemShut {NoStop}%
\bibitem [{\citenamefont {Zhang}\ and\ \citenamefont {Vishwanath}(2025)}]{JC_ashvin}%
  \BibitemOpen
  \bibfield  {author} {\bibinfo {author} {\bibfnamefont {Y.-H.}\ \bibnamefont {Zhang}}\ and\ \bibinfo {author} {\bibfnamefont {A.}~\bibnamefont {Vishwanath}},\ }\href {https://doi.org/10.36471/JCCM_February_2025_02} {\bibinfo {title} {Chiral superconductivity in the flat bands of rhombohedral graphene}} (\bibinfo {year} {2025})\BibitemShut {NoStop}%
\bibitem [{\citenamefont {Koshino}\ and\ \citenamefont {McCann}(2009)}]{Koshino_2009}%
  \BibitemOpen
  \bibfield  {author} {\bibinfo {author} {\bibfnamefont {M.}~\bibnamefont {Koshino}}\ and\ \bibinfo {author} {\bibfnamefont {E.}~\bibnamefont {McCann}},\ }\bibfield  {title} {\bibinfo {title} {Trigonal warping and berry's phase $n\ensuremath{\pi}$ in abc-stacked multilayer graphene},\ }\href {https://doi.org/10.1103/PhysRevB.80.165409} {\bibfield  {journal} {\bibinfo  {journal} {Phys. Rev. B}\ }\textbf {\bibinfo {volume} {80}},\ \bibinfo {pages} {165409} (\bibinfo {year} {2009})}\BibitemShut {NoStop}%
\bibitem [{\citenamefont {Zhang}\ \emph {et~al.}(2010)\citenamefont {Zhang}, \citenamefont {Sahu}, \citenamefont {Min},\ and\ \citenamefont {MacDonald}}]{Zhang2010}%
  \BibitemOpen
  \bibfield  {author} {\bibinfo {author} {\bibfnamefont {F.}~\bibnamefont {Zhang}}, \bibinfo {author} {\bibfnamefont {B.}~\bibnamefont {Sahu}}, \bibinfo {author} {\bibfnamefont {H.}~\bibnamefont {Min}},\ and\ \bibinfo {author} {\bibfnamefont {A.~H.}\ \bibnamefont {MacDonald}},\ }\bibfield  {title} {\bibinfo {title} {Band structure of $abc$-stacked graphene trilayers},\ }\href {https://doi.org/10.1103/PhysRevB.82.035409} {\bibfield  {journal} {\bibinfo  {journal} {Phys. Rev. B}\ }\textbf {\bibinfo {volume} {82}},\ \bibinfo {pages} {035409} (\bibinfo {year} {2010})}\BibitemShut {NoStop}%
\bibitem [{\citenamefont {Yuan}\ and\ \citenamefont {Fu}(2022)}]{Yuan_2022}%
  \BibitemOpen
  \bibfield  {author} {\bibinfo {author} {\bibfnamefont {N.~F.~Q.}\ \bibnamefont {Yuan}}\ and\ \bibinfo {author} {\bibfnamefont {L.}~\bibnamefont {Fu}},\ }\bibfield  {title} {\bibinfo {title} {Supercurrent diode effect and finite-momentum superconductors},\ }\href {https://doi.org/10.1073/pnas.2119548119} {\bibfield  {journal} {\bibinfo  {journal} {Proceedings of the National Academy of Sciences}\ }\textbf {\bibinfo {volume} {119}},\ \bibinfo {pages} {e2119548119} (\bibinfo {year} {2022})},\ \Eprint {https://arxiv.org/abs/https://www.pnas.org/doi/pdf/10.1073/pnas.2119548119} {https://www.pnas.org/doi/pdf/10.1073/pnas.2119548119} \BibitemShut {NoStop}%
\bibitem [{sup()}]{supplementary}%
  \BibitemOpen
  \href@noop {} {}\bibinfo {note} {See Supplementary Material at url ... for a detailed discussion of the Ginzburg-Landau free energy, the VAL order parameter, as well as the microscopic calculations of the free energy and the Bogoliubov-de Gennes continuum model.}\BibitemShut {Stop}%
\bibitem [{\citenamefont {Agterberg}\ and\ \citenamefont {Tsunetsugu}(2008)}]{Agterberg_2008}%
  \BibitemOpen
  \bibfield  {author} {\bibinfo {author} {\bibfnamefont {D.~F.}\ \bibnamefont {Agterberg}}\ and\ \bibinfo {author} {\bibfnamefont {H.}~\bibnamefont {Tsunetsugu}},\ }\bibfield  {title} {\bibinfo {title} {Dislocations and vortices in pair-density-wave superconductors},\ }\href {https://doi.org/10.1038/nphys999} {\bibfield  {journal} {\bibinfo  {journal} {Nature Physics}\ }\textbf {\bibinfo {volume} {4}},\ \bibinfo {pages} {639} (\bibinfo {year} {2008})}\BibitemShut {NoStop}%
\bibitem [{\citenamefont {Berg}\ \emph {et~al.}(2009)\citenamefont {Berg}, \citenamefont {Fradkin},\ and\ \citenamefont {Kivelson}}]{Berg_2009}%
  \BibitemOpen
  \bibfield  {author} {\bibinfo {author} {\bibfnamefont {E.}~\bibnamefont {Berg}}, \bibinfo {author} {\bibfnamefont {E.}~\bibnamefont {Fradkin}},\ and\ \bibinfo {author} {\bibfnamefont {S.~A.}\ \bibnamefont {Kivelson}},\ }\bibfield  {title} {\bibinfo {title} {Charge-4e superconductivity from pair-density-wave order in certain high-temperature superconductors},\ }\href {https://doi.org/10.1038/nphys1389} {\bibfield  {journal} {\bibinfo  {journal} {Nature Physics}\ }\textbf {\bibinfo {volume} {5}},\ \bibinfo {pages} {830} (\bibinfo {year} {2009})}\BibitemShut {NoStop}%
\bibitem [{\citenamefont {Agterberg}\ \emph {et~al.}(2020)\citenamefont {Agterberg}, \citenamefont {Davis}, \citenamefont {Edkins}, \citenamefont {Fradkin}, \citenamefont {Van~Harlingen}, \citenamefont {Kivelson}, \citenamefont {Lee}, \citenamefont {Radzihovsky}, \citenamefont {Tranquada},\ and\ \citenamefont {Wang}}]{Agterberg_2020_review}%
  \BibitemOpen
  \bibfield  {author} {\bibinfo {author} {\bibfnamefont {D.~F.}\ \bibnamefont {Agterberg}}, \bibinfo {author} {\bibfnamefont {J.~S.}\ \bibnamefont {Davis}}, \bibinfo {author} {\bibfnamefont {S.~D.}\ \bibnamefont {Edkins}}, \bibinfo {author} {\bibfnamefont {E.}~\bibnamefont {Fradkin}}, \bibinfo {author} {\bibfnamefont {D.~J.}\ \bibnamefont {Van~Harlingen}}, \bibinfo {author} {\bibfnamefont {S.~A.}\ \bibnamefont {Kivelson}}, \bibinfo {author} {\bibfnamefont {P.~A.}\ \bibnamefont {Lee}}, \bibinfo {author} {\bibfnamefont {L.}~\bibnamefont {Radzihovsky}}, \bibinfo {author} {\bibfnamefont {J.~M.}\ \bibnamefont {Tranquada}},\ and\ \bibinfo {author} {\bibfnamefont {Y.}~\bibnamefont {Wang}},\ }\bibfield  {title} {\bibinfo {title} {The physics of pair-density waves: Cuprate superconductors and beyond},\ }\href {https://doi.org/https://doi.org/10.1146/annurev-conmatphys-031119-050711} {\bibfield  {journal} {\bibinfo  {journal} {Annual Review of Condensed Matter Physics}\ }\textbf {\bibinfo {volume} {11}},\ \bibinfo
  {pages} {231} (\bibinfo {year} {2020})}\BibitemShut {NoStop}%
\bibitem [{\citenamefont {Yin}\ \emph {et~al.}(2022)\citenamefont {Yin}, \citenamefont {Lian},\ and\ \citenamefont {Hasan}}]{Yin_2022_review}%
  \BibitemOpen
  \bibfield  {author} {\bibinfo {author} {\bibfnamefont {J.-X.}\ \bibnamefont {Yin}}, \bibinfo {author} {\bibfnamefont {B.}~\bibnamefont {Lian}},\ and\ \bibinfo {author} {\bibfnamefont {M.~Z.}\ \bibnamefont {Hasan}},\ }\bibfield  {title} {\bibinfo {title} {Topological kagome magnets and superconductors},\ }\href {https://doi.org/10.1038/s41586-022-05516-0} {\bibfield  {journal} {\bibinfo  {journal} {Nature}\ }\textbf {\bibinfo {volume} {612}},\ \bibinfo {pages} {647} (\bibinfo {year} {2022})}\BibitemShut {NoStop}%
\bibitem [{\citenamefont {Neupert}\ \emph {et~al.}(2022)\citenamefont {Neupert}, \citenamefont {Denner}, \citenamefont {Yin}, \citenamefont {Thomale},\ and\ \citenamefont {Hasan}}]{Neupert_2022_review}%
  \BibitemOpen
  \bibfield  {author} {\bibinfo {author} {\bibfnamefont {T.}~\bibnamefont {Neupert}}, \bibinfo {author} {\bibfnamefont {M.~M.}\ \bibnamefont {Denner}}, \bibinfo {author} {\bibfnamefont {J.-X.}\ \bibnamefont {Yin}}, \bibinfo {author} {\bibfnamefont {R.}~\bibnamefont {Thomale}},\ and\ \bibinfo {author} {\bibfnamefont {M.~Z.}\ \bibnamefont {Hasan}},\ }\bibfield  {title} {\bibinfo {title} {Charge order and superconductivity in kagome materials},\ }\href {https://doi.org/10.1038/s41567-021-01404-y} {\bibfield  {journal} {\bibinfo  {journal} {Nature Physics}\ }\textbf {\bibinfo {volume} {18}},\ \bibinfo {pages} {137} (\bibinfo {year} {2022})}\BibitemShut {NoStop}%
\bibitem [{\citenamefont {Deng}\ \emph {et~al.}(2024)\citenamefont {Deng}, \citenamefont {Qin}, \citenamefont {Liu}, \citenamefont {Yang}, \citenamefont {Fu}, \citenamefont {Zhang}, \citenamefont {Wu}, \citenamefont {Wang}, \citenamefont {Shi}, \citenamefont {Liu}, \citenamefont {Liu}, \citenamefont {Yan}, \citenamefont {Song}, \citenamefont {Xu}, \citenamefont {Zhao}, \citenamefont {Yi}, \citenamefont {Xu}, \citenamefont {Hohmann}, \citenamefont {Holb{\ae}k}, \citenamefont {D{\"u}rrnagel}, \citenamefont {Zhou}, \citenamefont {Chang}, \citenamefont {Yao}, \citenamefont {Wang}, \citenamefont {Guguchia}, \citenamefont {Neupert}, \citenamefont {Thomale}, \citenamefont {Fischer},\ and\ \citenamefont {Yin}}]{Deng_2024}%
  \BibitemOpen
  \bibfield  {author} {\bibinfo {author} {\bibfnamefont {H.}~\bibnamefont {Deng}}, \bibinfo {author} {\bibfnamefont {H.}~\bibnamefont {Qin}}, \bibinfo {author} {\bibfnamefont {G.}~\bibnamefont {Liu}}, \bibinfo {author} {\bibfnamefont {T.}~\bibnamefont {Yang}}, \bibinfo {author} {\bibfnamefont {R.}~\bibnamefont {Fu}}, \bibinfo {author} {\bibfnamefont {Z.}~\bibnamefont {Zhang}}, \bibinfo {author} {\bibfnamefont {X.}~\bibnamefont {Wu}}, \bibinfo {author} {\bibfnamefont {Z.}~\bibnamefont {Wang}}, \bibinfo {author} {\bibfnamefont {Y.}~\bibnamefont {Shi}}, \bibinfo {author} {\bibfnamefont {J.}~\bibnamefont {Liu}}, \bibinfo {author} {\bibfnamefont {H.}~\bibnamefont {Liu}}, \bibinfo {author} {\bibfnamefont {X.-Y.}\ \bibnamefont {Yan}}, \bibinfo {author} {\bibfnamefont {W.}~\bibnamefont {Song}}, \bibinfo {author} {\bibfnamefont {X.}~\bibnamefont {Xu}}, \bibinfo {author} {\bibfnamefont {Y.}~\bibnamefont {Zhao}}, \bibinfo {author} {\bibfnamefont {M.}~\bibnamefont {Yi}}, \bibinfo {author} {\bibfnamefont {G.}~\bibnamefont
  {Xu}}, \bibinfo {author} {\bibfnamefont {H.}~\bibnamefont {Hohmann}}, \bibinfo {author} {\bibfnamefont {S.~C.}\ \bibnamefont {Holb{\ae}k}}, \bibinfo {author} {\bibfnamefont {M.}~\bibnamefont {D{\"u}rrnagel}}, \bibinfo {author} {\bibfnamefont {S.}~\bibnamefont {Zhou}}, \bibinfo {author} {\bibfnamefont {G.}~\bibnamefont {Chang}}, \bibinfo {author} {\bibfnamefont {Y.}~\bibnamefont {Yao}}, \bibinfo {author} {\bibfnamefont {Q.}~\bibnamefont {Wang}}, \bibinfo {author} {\bibfnamefont {Z.}~\bibnamefont {Guguchia}}, \bibinfo {author} {\bibfnamefont {T.}~\bibnamefont {Neupert}}, \bibinfo {author} {\bibfnamefont {R.}~\bibnamefont {Thomale}}, \bibinfo {author} {\bibfnamefont {M.~H.}\ \bibnamefont {Fischer}},\ and\ \bibinfo {author} {\bibfnamefont {J.-X.}\ \bibnamefont {Yin}},\ }\bibfield  {title} {\bibinfo {title} {Chiral kagome superconductivity modulations with residual fermi arcs},\ }\href {https://doi.org/10.1038/s41586-024-07798-y} {\bibfield  {journal} {\bibinfo  {journal} {Nature}\ }\textbf {\bibinfo {volume}
  {632}},\ \bibinfo {pages} {775} (\bibinfo {year} {2024})}\BibitemShut {NoStop}%
\bibitem [{\citenamefont {Agterberg}\ \emph {et~al.}(2011)\citenamefont {Agterberg}, \citenamefont {Geracie},\ and\ \citenamefont {Tsunetsugu}}]{Agterberg_2011}%
  \BibitemOpen
  \bibfield  {author} {\bibinfo {author} {\bibfnamefont {D.~F.}\ \bibnamefont {Agterberg}}, \bibinfo {author} {\bibfnamefont {M.}~\bibnamefont {Geracie}},\ and\ \bibinfo {author} {\bibfnamefont {H.}~\bibnamefont {Tsunetsugu}},\ }\bibfield  {title} {\bibinfo {title} {Conventional and charge-six superfluids from melting hexagonal fulde-ferrell-larkin-ovchinnikov phases in two dimensions},\ }\href {https://doi.org/10.1103/PhysRevB.84.014513} {\bibfield  {journal} {\bibinfo  {journal} {Phys. Rev. B}\ }\textbf {\bibinfo {volume} {84}},\ \bibinfo {pages} {014513} (\bibinfo {year} {2011})}\BibitemShut {NoStop}%
\bibitem [{\citenamefont {Vasyukov}\ \emph {et~al.}(2013)\citenamefont {Vasyukov}, \citenamefont {Anahory}, \citenamefont {Embon}, \citenamefont {Halbertal}, \citenamefont {Cuppens}, \citenamefont {Neeman}, \citenamefont {Finkler}, \citenamefont {Segev}, \citenamefont {Myasoedov}, \citenamefont {Rappaport}, \citenamefont {Huber},\ and\ \citenamefont {Zeldov}}]{Vasyukov_2013}%
  \BibitemOpen
  \bibfield  {author} {\bibinfo {author} {\bibfnamefont {D.}~\bibnamefont {Vasyukov}}, \bibinfo {author} {\bibfnamefont {Y.}~\bibnamefont {Anahory}}, \bibinfo {author} {\bibfnamefont {L.}~\bibnamefont {Embon}}, \bibinfo {author} {\bibfnamefont {D.}~\bibnamefont {Halbertal}}, \bibinfo {author} {\bibfnamefont {J.}~\bibnamefont {Cuppens}}, \bibinfo {author} {\bibfnamefont {L.}~\bibnamefont {Neeman}}, \bibinfo {author} {\bibfnamefont {A.}~\bibnamefont {Finkler}}, \bibinfo {author} {\bibfnamefont {Y.}~\bibnamefont {Segev}}, \bibinfo {author} {\bibfnamefont {Y.}~\bibnamefont {Myasoedov}}, \bibinfo {author} {\bibfnamefont {M.~L.}\ \bibnamefont {Rappaport}}, \bibinfo {author} {\bibfnamefont {M.~E.}\ \bibnamefont {Huber}},\ and\ \bibinfo {author} {\bibfnamefont {E.}~\bibnamefont {Zeldov}},\ }\bibfield  {title} {\bibinfo {title} {A scanning superconducting quantum interference device with single electron spin sensitivity},\ }\href {https://doi.org/10.1038/nnano.2013.169} {\bibfield  {journal} {\bibinfo  {journal}
  {Nature Nanotechnology}\ }\textbf {\bibinfo {volume} {8}},\ \bibinfo {pages} {639} (\bibinfo {year} {2013})}\BibitemShut {NoStop}%
\bibitem [{\citenamefont {Uri}\ \emph {et~al.}(2016)\citenamefont {Uri}, \citenamefont {Meltzer}, \citenamefont {Anahory}, \citenamefont {Embon}, \citenamefont {Lachman}, \citenamefont {Halbertal}, \citenamefont {HR}, \citenamefont {Myasoedov}, \citenamefont {Huber}, \citenamefont {Young},\ and\ \citenamefont {Zeldov}}]{Uri_2016}%
  \BibitemOpen
  \bibfield  {author} {\bibinfo {author} {\bibfnamefont {A.}~\bibnamefont {Uri}}, \bibinfo {author} {\bibfnamefont {A.~Y.}\ \bibnamefont {Meltzer}}, \bibinfo {author} {\bibfnamefont {Y.}~\bibnamefont {Anahory}}, \bibinfo {author} {\bibfnamefont {L.}~\bibnamefont {Embon}}, \bibinfo {author} {\bibfnamefont {E.~O.}\ \bibnamefont {Lachman}}, \bibinfo {author} {\bibfnamefont {D.}~\bibnamefont {Halbertal}}, \bibinfo {author} {\bibfnamefont {N.}~\bibnamefont {HR}}, \bibinfo {author} {\bibfnamefont {Y.}~\bibnamefont {Myasoedov}}, \bibinfo {author} {\bibfnamefont {M.~E.}\ \bibnamefont {Huber}}, \bibinfo {author} {\bibfnamefont {A.~F.}\ \bibnamefont {Young}},\ and\ \bibinfo {author} {\bibfnamefont {E.}~\bibnamefont {Zeldov}},\ }\bibfield  {title} {\bibinfo {title} {Electrically tunable multiterminal squid-on-tip},\ }\href {https://doi.org/10.1021/acs.nanolett.6b02841} {\bibfield  {journal} {\bibinfo  {journal} {Nano Letters}\ }\textbf {\bibinfo {volume} {16}},\ \bibinfo {pages} {6910} (\bibinfo {year}
  {2016})}\BibitemShut {NoStop}%
\bibitem [{\citenamefont {Gaggioli}\ \emph {et~al.}(2024)\citenamefont {Gaggioli}, \citenamefont {Blatter}, \citenamefont {Novoselov},\ and\ \citenamefont {Geshkenbein}}]{Gaggioli_2024}%
  \BibitemOpen
  \bibfield  {author} {\bibinfo {author} {\bibfnamefont {F.}~\bibnamefont {Gaggioli}}, \bibinfo {author} {\bibfnamefont {G.}~\bibnamefont {Blatter}}, \bibinfo {author} {\bibfnamefont {K.~S.}\ \bibnamefont {Novoselov}},\ and\ \bibinfo {author} {\bibfnamefont {V.~B.}\ \bibnamefont {Geshkenbein}},\ }\bibfield  {title} {\bibinfo {title} {Superconductivity in atomically thin films: Two-dimensional critical state model},\ }\href {https://doi.org/10.1103/PhysRevResearch.6.023190} {\bibfield  {journal} {\bibinfo  {journal} {Phys. Rev. Res.}\ }\textbf {\bibinfo {volume} {6}},\ \bibinfo {pages} {023190} (\bibinfo {year} {2024})}\BibitemShut {NoStop}%
\bibitem [{\citenamefont {Blatter}\ \emph {et~al.}(1994)\citenamefont {Blatter}, \citenamefont {Feigel'man}, \citenamefont {Geshkenbein}, \citenamefont {Larkin},\ and\ \citenamefont {Vinokur}}]{Blatter_1994}%
  \BibitemOpen
  \bibfield  {author} {\bibinfo {author} {\bibfnamefont {G.}~\bibnamefont {Blatter}}, \bibinfo {author} {\bibfnamefont {M.~V.}\ \bibnamefont {Feigel'man}}, \bibinfo {author} {\bibfnamefont {V.~B.}\ \bibnamefont {Geshkenbein}}, \bibinfo {author} {\bibfnamefont {A.~I.}\ \bibnamefont {Larkin}},\ and\ \bibinfo {author} {\bibfnamefont {V.~M.}\ \bibnamefont {Vinokur}},\ }\bibfield  {title} {\bibinfo {title} {Vortices in high-temperature superconductors},\ }\href {https://doi.org/10.1103/RevModPhys.66.1125} {\bibfield  {journal} {\bibinfo  {journal} {Rev. Mod. Phys.}\ }\textbf {\bibinfo {volume} {66}},\ \bibinfo {pages} {1125} (\bibinfo {year} {1994})}\BibitemShut {NoStop}%
\bibitem [{\citenamefont {Stephen}\ and\ \citenamefont {Bardeen}(1965)}]{Stephen_1965}%
  \BibitemOpen
  \bibfield  {author} {\bibinfo {author} {\bibfnamefont {M.~J.}\ \bibnamefont {Stephen}}\ and\ \bibinfo {author} {\bibfnamefont {J.}~\bibnamefont {Bardeen}},\ }\bibfield  {title} {\bibinfo {title} {Viscosity of type-ii superconductors},\ }\href {https://doi.org/10.1103/PhysRevLett.14.112} {\bibfield  {journal} {\bibinfo  {journal} {Phys. Rev. Lett.}\ }\textbf {\bibinfo {volume} {14}},\ \bibinfo {pages} {112} (\bibinfo {year} {1965})}\BibitemShut {NoStop}%
\bibitem [{\citenamefont {Franz}\ and\ \citenamefont {Te\ifmmode \check{s}\else \v{s}\fi{}anovi\ifmmode~\acute{c}\else \'{c}\fi{}}(2000)}]{Franz_2000}%
  \BibitemOpen
  \bibfield  {author} {\bibinfo {author} {\bibfnamefont {M.}~\bibnamefont {Franz}}\ and\ \bibinfo {author} {\bibfnamefont {Z.}~\bibnamefont {Te\ifmmode \check{s}\else \v{s}\fi{}anovi\ifmmode~\acute{c}\else \'{c}\fi{}}},\ }\bibfield  {title} {\bibinfo {title} {Quasiparticles in the vortex lattice of unconventional superconductors: Bloch waves or landau levels?},\ }\href {https://doi.org/10.1103/PhysRevLett.84.554} {\bibfield  {journal} {\bibinfo  {journal} {Phys. Rev. Lett.}\ }\textbf {\bibinfo {volume} {84}},\ \bibinfo {pages} {554} (\bibinfo {year} {2000})}\BibitemShut {NoStop}%
\bibitem [{\citenamefont {Vafek}\ \emph {et~al.}(2001)\citenamefont {Vafek}, \citenamefont {Melikyan},\ and\ \citenamefont {Te\ifmmode \check{s}\else \v{s}\fi{}anovi\ifmmode~\acute{c}\else \'{c}\fi{}}}]{Vafek_2001}%
  \BibitemOpen
  \bibfield  {author} {\bibinfo {author} {\bibfnamefont {O.}~\bibnamefont {Vafek}}, \bibinfo {author} {\bibfnamefont {A.}~\bibnamefont {Melikyan}},\ and\ \bibinfo {author} {\bibfnamefont {Z.}~\bibnamefont {Te\ifmmode \check{s}\else \v{s}\fi{}anovi\ifmmode~\acute{c}\else \'{c}\fi{}}},\ }\bibfield  {title} {\bibinfo {title} {Quasiparticle hall transport of d-wave superconductors in the vortex state},\ }\href {https://doi.org/10.1103/PhysRevB.64.224508} {\bibfield  {journal} {\bibinfo  {journal} {Phys. Rev. B}\ }\textbf {\bibinfo {volume} {64}},\ \bibinfo {pages} {224508} (\bibinfo {year} {2001})}\BibitemShut {NoStop}%
\bibitem [{\citenamefont {Chiu}\ \emph {et~al.}(2015)\citenamefont {Chiu}, \citenamefont {Pikulin},\ and\ \citenamefont {Franz}}]{Chiu_2015}%
  \BibitemOpen
  \bibfield  {author} {\bibinfo {author} {\bibfnamefont {C.-K.}\ \bibnamefont {Chiu}}, \bibinfo {author} {\bibfnamefont {D.~I.}\ \bibnamefont {Pikulin}},\ and\ \bibinfo {author} {\bibfnamefont {M.}~\bibnamefont {Franz}},\ }\bibfield  {title} {\bibinfo {title} {Strongly interacting majorana fermions},\ }\href {https://doi.org/10.1103/PhysRevB.91.165402} {\bibfield  {journal} {\bibinfo  {journal} {Phys. Rev. B}\ }\textbf {\bibinfo {volume} {91}},\ \bibinfo {pages} {165402} (\bibinfo {year} {2015})}\BibitemShut {NoStop}%
\bibitem [{\citenamefont {Liu}\ and\ \citenamefont {Franz}(2015)}]{Liu_2015}%
  \BibitemOpen
  \bibfield  {author} {\bibinfo {author} {\bibfnamefont {T.}~\bibnamefont {Liu}}\ and\ \bibinfo {author} {\bibfnamefont {M.}~\bibnamefont {Franz}},\ }\bibfield  {title} {\bibinfo {title} {Electronic structure of topological superconductors in the presence of a vortex lattice},\ }\href {https://doi.org/10.1103/PhysRevB.92.134519} {\bibfield  {journal} {\bibinfo  {journal} {Phys. Rev. B}\ }\textbf {\bibinfo {volume} {92}},\ \bibinfo {pages} {134519} (\bibinfo {year} {2015})}\BibitemShut {NoStop}%
\bibitem [{\citenamefont {Pu}\ \emph {et~al.}(2024)\citenamefont {Pu}, \citenamefont {Sau},\ and\ \citenamefont {Zhang}}]{Pu_2024}%
  \BibitemOpen
  \bibfield  {author} {\bibinfo {author} {\bibfnamefont {S.}~\bibnamefont {Pu}}, \bibinfo {author} {\bibfnamefont {J.~D.}\ \bibnamefont {Sau}},\ and\ \bibinfo {author} {\bibfnamefont {R.-X.}\ \bibnamefont {Zhang}},\ }\href {https://arxiv.org/abs/2402.18627} {\bibinfo {title} {Topologically protected emergent fermi surface in an abrikosov vortex lattice}} (\bibinfo {year} {2024}),\ \Eprint {https://arxiv.org/abs/2402.18627} {arXiv:2402.18627 [cond-mat.supr-con]} \BibitemShut {NoStop}%
\bibitem [{\citenamefont {Brydon}\ \emph {et~al.}(2018)\citenamefont {Brydon}, \citenamefont {Agterberg}, \citenamefont {Menke},\ and\ \citenamefont {Timm}}]{Brydon_2018}%
  \BibitemOpen
  \bibfield  {author} {\bibinfo {author} {\bibfnamefont {P.~M.~R.}\ \bibnamefont {Brydon}}, \bibinfo {author} {\bibfnamefont {D.~F.}\ \bibnamefont {Agterberg}}, \bibinfo {author} {\bibfnamefont {H.}~\bibnamefont {Menke}},\ and\ \bibinfo {author} {\bibfnamefont {C.}~\bibnamefont {Timm}},\ }\bibfield  {title} {\bibinfo {title} {Bogoliubov fermi surfaces: General theory, magnetic order, and topology},\ }\href {https://doi.org/10.1103/PhysRevB.98.224509} {\bibfield  {journal} {\bibinfo  {journal} {Phys. Rev. B}\ }\textbf {\bibinfo {volume} {98}},\ \bibinfo {pages} {224509} (\bibinfo {year} {2018})}\BibitemShut {NoStop}%
\bibitem [{\citenamefont {Yuan}\ and\ \citenamefont {Fu}(2018)}]{Yuan_2018}%
  \BibitemOpen
  \bibfield  {author} {\bibinfo {author} {\bibfnamefont {N.~F.~Q.}\ \bibnamefont {Yuan}}\ and\ \bibinfo {author} {\bibfnamefont {L.}~\bibnamefont {Fu}},\ }\bibfield  {title} {\bibinfo {title} {Zeeman-induced gapless superconductivity with a partial fermi surface},\ }\href {https://doi.org/10.1103/PhysRevB.97.115139} {\bibfield  {journal} {\bibinfo  {journal} {Phys. Rev. B}\ }\textbf {\bibinfo {volume} {97}},\ \bibinfo {pages} {115139} (\bibinfo {year} {2018})}\BibitemShut {NoStop}%
\bibitem [{\citenamefont {Kraus}\ \emph {et~al.}(2009)\citenamefont {Kraus}, \citenamefont {Auerbach}, \citenamefont {Fertig},\ and\ \citenamefont {Simon}}]{Kraus_2009}%
  \BibitemOpen
  \bibfield  {author} {\bibinfo {author} {\bibfnamefont {Y.~E.}\ \bibnamefont {Kraus}}, \bibinfo {author} {\bibfnamefont {A.}~\bibnamefont {Auerbach}}, \bibinfo {author} {\bibfnamefont {H.~A.}\ \bibnamefont {Fertig}},\ and\ \bibinfo {author} {\bibfnamefont {S.~H.}\ \bibnamefont {Simon}},\ }\bibfield  {title} {\bibinfo {title} {Majorana fermions of a two-dimensional ${p}_{x}+i{p}_{y}$ superconductor},\ }\href {https://doi.org/10.1103/PhysRevB.79.134515} {\bibfield  {journal} {\bibinfo  {journal} {Phys. Rev. B}\ }\textbf {\bibinfo {volume} {79}},\ \bibinfo {pages} {134515} (\bibinfo {year} {2009})}\BibitemShut {NoStop}%
\bibitem [{\citenamefont {Haldane}(1988)}]{Haldane_1988}%
  \BibitemOpen
  \bibfield  {author} {\bibinfo {author} {\bibfnamefont {F.~D.~M.}\ \bibnamefont {Haldane}},\ }\bibfield  {title} {\bibinfo {title} {Model for a quantum hall effect without landau levels: Condensed-matter realization of the "parity anomaly"},\ }\href {https://doi.org/10.1103/PhysRevLett.61.2015} {\bibfield  {journal} {\bibinfo  {journal} {Phys. Rev. Lett.}\ }\textbf {\bibinfo {volume} {61}},\ \bibinfo {pages} {2015} (\bibinfo {year} {1988})}\BibitemShut {NoStop}%
\bibitem [{\citenamefont {Kitaev}(2006)}]{Kitaev_2006}%
  \BibitemOpen
  \bibfield  {author} {\bibinfo {author} {\bibfnamefont {A.}~\bibnamefont {Kitaev}},\ }\bibfield  {title} {\bibinfo {title} {Anyons in an exactly solved model and beyond},\ }\href {https://doi.org/https://doi.org/10.1016/j.aop.2005.10.005} {\bibfield  {journal} {\bibinfo  {journal} {Annals of Physics}\ }\textbf {\bibinfo {volume} {321}},\ \bibinfo {pages} {2} (\bibinfo {year} {2006})},\ \bibinfo {note} {january Special Issue}\BibitemShut {NoStop}%
\bibitem [{\citenamefont {Pomeranchuk}(1950)}]{Pomeranchuk_1950}%
  \BibitemOpen
  \bibfield  {author} {\bibinfo {author} {\bibfnamefont {I.~Y.}\ \bibnamefont {Pomeranchuk}},\ }\bibfield  {title} {\bibinfo {title} {On the theory of he3},\ }\href@noop {} {\bibfield  {journal} {\bibinfo  {journal} {Zh. Eksp. i Teor. Fiz.}\ }\textbf {\bibinfo {volume} {20}},\ \bibinfo {pages} {919} (\bibinfo {year} {1950})}\BibitemShut {NoStop}%
\bibitem [{\citenamefont {Rozen}\ \emph {et~al.}(2021)\citenamefont {Rozen}, \citenamefont {Park}, \citenamefont {Zondiner}, \citenamefont {Cao}, \citenamefont {Rodan-Legrain}, \citenamefont {Taniguchi}, \citenamefont {Watanabe}, \citenamefont {Oreg}, \citenamefont {Stern}, \citenamefont {Berg}, \citenamefont {Jarillo-Herrero},\ and\ \citenamefont {Ilani}}]{Rozen_2021}%
  \BibitemOpen
  \bibfield  {author} {\bibinfo {author} {\bibfnamefont {A.}~\bibnamefont {Rozen}}, \bibinfo {author} {\bibfnamefont {J.~M.}\ \bibnamefont {Park}}, \bibinfo {author} {\bibfnamefont {U.}~\bibnamefont {Zondiner}}, \bibinfo {author} {\bibfnamefont {Y.}~\bibnamefont {Cao}}, \bibinfo {author} {\bibfnamefont {D.}~\bibnamefont {Rodan-Legrain}}, \bibinfo {author} {\bibfnamefont {T.}~\bibnamefont {Taniguchi}}, \bibinfo {author} {\bibfnamefont {K.}~\bibnamefont {Watanabe}}, \bibinfo {author} {\bibfnamefont {Y.}~\bibnamefont {Oreg}}, \bibinfo {author} {\bibfnamefont {A.}~\bibnamefont {Stern}}, \bibinfo {author} {\bibfnamefont {E.}~\bibnamefont {Berg}}, \bibinfo {author} {\bibfnamefont {P.}~\bibnamefont {Jarillo-Herrero}},\ and\ \bibinfo {author} {\bibfnamefont {S.}~\bibnamefont {Ilani}},\ }\bibfield  {title} {\bibinfo {title} {Entropic evidence for a pomeranchuk effect in magic-angle graphene},\ }\href {https://doi.org/10.1038/s41586-021-03319-3} {\bibfield  {journal} {\bibinfo  {journal} {Nature}\ }\textbf {\bibinfo
  {volume} {592}},\ \bibinfo {pages} {214} (\bibinfo {year} {2021})}\BibitemShut {NoStop}%
\bibitem [{\citenamefont {Saito}\ \emph {et~al.}(2021)\citenamefont {Saito}, \citenamefont {Yang}, \citenamefont {Ge}, \citenamefont {Liu}, \citenamefont {Taniguchi}, \citenamefont {Watanabe}, \citenamefont {Li}, \citenamefont {Berg},\ and\ \citenamefont {Young}}]{Saito_2021}%
  \BibitemOpen
  \bibfield  {author} {\bibinfo {author} {\bibfnamefont {Y.}~\bibnamefont {Saito}}, \bibinfo {author} {\bibfnamefont {F.}~\bibnamefont {Yang}}, \bibinfo {author} {\bibfnamefont {J.}~\bibnamefont {Ge}}, \bibinfo {author} {\bibfnamefont {X.}~\bibnamefont {Liu}}, \bibinfo {author} {\bibfnamefont {T.}~\bibnamefont {Taniguchi}}, \bibinfo {author} {\bibfnamefont {K.}~\bibnamefont {Watanabe}}, \bibinfo {author} {\bibfnamefont {J.~I.~A.}\ \bibnamefont {Li}}, \bibinfo {author} {\bibfnamefont {E.}~\bibnamefont {Berg}},\ and\ \bibinfo {author} {\bibfnamefont {A.~F.}\ \bibnamefont {Young}},\ }\bibfield  {title} {\bibinfo {title} {Isospin pomeranchuk effect in twisted bilayer graphene},\ }\href {https://doi.org/10.1038/s41586-021-03409-2} {\bibfield  {journal} {\bibinfo  {journal} {Nature}\ }\textbf {\bibinfo {volume} {592}},\ \bibinfo {pages} {220} (\bibinfo {year} {2021})}\BibitemShut {NoStop}%
\bibitem [{\citenamefont {Zhang}(1993)}]{Zhang_1993}%
  \BibitemOpen
  \bibfield  {author} {\bibinfo {author} {\bibfnamefont {S.-C.}\ \bibnamefont {Zhang}},\ }\bibfield  {title} {\bibinfo {title} {Vortex-antivortex lattice in superfluid films},\ }\href {https://doi.org/10.1103/PhysRevLett.71.2142} {\bibfield  {journal} {\bibinfo  {journal} {Phys. Rev. Lett.}\ }\textbf {\bibinfo {volume} {71}},\ \bibinfo {pages} {2142} (\bibinfo {year} {1993})}\BibitemShut {NoStop}%
\bibitem [{\citenamefont {Gabay}\ and\ \citenamefont {Kapitulnik}(1993)}]{Gabay_1993}%
  \BibitemOpen
  \bibfield  {author} {\bibinfo {author} {\bibfnamefont {M.}~\bibnamefont {Gabay}}\ and\ \bibinfo {author} {\bibfnamefont {A.}~\bibnamefont {Kapitulnik}},\ }\bibfield  {title} {\bibinfo {title} {Vortex-antivortex crystallization in thin superconducting and superfluid films},\ }\href {https://doi.org/10.1103/PhysRevLett.71.2138} {\bibfield  {journal} {\bibinfo  {journal} {Phys. Rev. Lett.}\ }\textbf {\bibinfo {volume} {71}},\ \bibinfo {pages} {2138} (\bibinfo {year} {1993})}\BibitemShut {NoStop}%
\bibitem [{\citenamefont {Christos}\ \emph {et~al.}(2025)\citenamefont {Christos}, \citenamefont {Bonetti},\ and\ \citenamefont {Scheurer}}]{Christos_2025}%
  \BibitemOpen
  \bibfield  {author} {\bibinfo {author} {\bibfnamefont {M.}~\bibnamefont {Christos}}, \bibinfo {author} {\bibfnamefont {P.~M.}\ \bibnamefont {Bonetti}},\ and\ \bibinfo {author} {\bibfnamefont {M.~S.}\ \bibnamefont {Scheurer}},\ }\href {https://arxiv.org/abs/2503.15471} {\bibinfo {title} {Finite-momentum pairing and superlattice superconductivity in valley-imbalanced rhombohedral graphene}} (\bibinfo {year} {2025}),\ \Eprint {https://arxiv.org/abs/2503.15471} {arXiv:2503.15471 [cond-mat.str-el]} \BibitemShut {NoStop}%
\bibitem [{\citenamefont {Gil}\ and\ \citenamefont {Berg}(2025)}]{Gil_2025}%
  \BibitemOpen
  \bibfield  {author} {\bibinfo {author} {\bibfnamefont {A.}~\bibnamefont {Gil}}\ and\ \bibinfo {author} {\bibfnamefont {E.}~\bibnamefont {Berg}},\ }\href {https://arxiv.org/abs/2504.19321} {\bibinfo {title} {Charge and pair density waves in a spin and valley-polarized system at a van-hove singularity}} (\bibinfo {year} {2025}),\ \Eprint {https://arxiv.org/abs/2504.19321} {arXiv:2504.19321 [cond-mat.str-el]} \BibitemShut {NoStop}%
\bibitem [{\citenamefont {Jung}\ and\ \citenamefont {MacDonald}(2013)}]{Jung2013}%
  \BibitemOpen
  \bibfield  {author} {\bibinfo {author} {\bibfnamefont {J.}~\bibnamefont {Jung}}\ and\ \bibinfo {author} {\bibfnamefont {A.~H.}\ \bibnamefont {MacDonald}},\ }\bibfield  {title} {\bibinfo {title} {Gapped broken symmetry states in abc-stacked trilayer graphene},\ }\href {https://doi.org/10.1103/PhysRevB.88.075408} {\bibfield  {journal} {\bibinfo  {journal} {Phys. Rev. B}\ }\textbf {\bibinfo {volume} {88}},\ \bibinfo {pages} {075408} (\bibinfo {year} {2013})}\BibitemShut {NoStop}%
\bibitem [{\citenamefont {Dong}\ \emph {et~al.}(2024)\citenamefont {Dong}, \citenamefont {Wang}, \citenamefont {Wang}, \citenamefont {Soejima}, \citenamefont {Zaletel}, \citenamefont {Vishwanath},\ and\ \citenamefont {Parker}}]{Dong_2024}%
  \BibitemOpen
  \bibfield  {author} {\bibinfo {author} {\bibfnamefont {J.}~\bibnamefont {Dong}}, \bibinfo {author} {\bibfnamefont {T.}~\bibnamefont {Wang}}, \bibinfo {author} {\bibfnamefont {T.}~\bibnamefont {Wang}}, \bibinfo {author} {\bibfnamefont {T.}~\bibnamefont {Soejima}}, \bibinfo {author} {\bibfnamefont {M.~P.}\ \bibnamefont {Zaletel}}, \bibinfo {author} {\bibfnamefont {A.}~\bibnamefont {Vishwanath}},\ and\ \bibinfo {author} {\bibfnamefont {D.~E.}\ \bibnamefont {Parker}},\ }\bibfield  {title} {\bibinfo {title} {Anomalous hall crystals in rhombohedral multilayer graphene. i. interaction-driven chern bands and fractional quantum hall states at zero magnetic field},\ }\bibfield  {journal} {\bibinfo  {journal} {Physical Review Letters}\ }\textbf {\bibinfo {volume} {133}},\ \href {https://doi.org/10.1103/physrevlett.133.206503} {10.1103/physrevlett.133.206503} (\bibinfo {year} {2024})\BibitemShut {NoStop}%
\end{thebibliography}%

\onecolumngrid
\newpage
\makeatletter 

\begin{center}
\textbf{\large Supplementary materials for: ``Spontaneous vortex-antivortex lattice and Majorana fermions in rhombohedral graphene ''} \\[10pt]
Filippo Gaggioli$^{1}$, Daniele Guerci$^{1}$ and Liang Fu$^{1}$ \\
\textit{$^1$Department of Physics, Massachusetts Institute of Technology, Cambridge, MA, USA}
\end{center}
\vspace{20pt}

\setcounter{figure}{0}
\setcounter{section}{0}
\setcounter{equation}{0}

\renewcommand{\thefigure}{S\@arabic\c@figure}
\makeatother

 \appendix

These supplementary materials contain the details of the Ginzburg-Landau theory as well as microscopic calculations supporting the results presented in the main text.
Sec.~\ref{app:scaling_analysis} and~\ref{app:scaling_analysis_1} contain details on the symmetry-constrainted expression of the particle-particle correlation function $K(\qb)$.  
In Sec.~\ref{app:order_parameter} we show that the order parameter of the VAL is unambiguously determined, i.e. any additional phase in the Fourier components $\psi_j=|\psi_c|\to|\psi_c|e^{i\varphi_j}$ is absorbed by a gauge transformation. 
Sec.~\ref{app:microscopic} and Sec.~\ref{app:GL_microscopic} provide microscopic results establishing a solid ground for our theory. The continuum model Hamiltonian for the VAL is discussed in Sec.~\ref{sec:BdG}.

\section{Scaling properties and phase diagram of the quadratic kernel $K(q)$} \label{app:scaling_analysis}

In this section we want to determine the phase diagram associated to the quadratic kernel
\begin{align}\label{eq:kernel_appendix}
    K(q) = \alpha(T) + \left(\hbar^2/2m^*\right)q^2  -\gamma\,q^3 + \eta q^4.
\end{align}
First, we want to reduce the parameter space by introducing the momentum scale $q_c = \sqrt{\hbar^2/2m^*\eta}$ and the dimensionless variable $t = q/q_c$, such that
\begin{align}\label{eq:rescaled_kernel}
     \Delta K(q) =  \eta\left(\frac{\hbar^2}{2m^*\eta}\right)^2t^2  + \gamma \left(\frac{\hbar^2}{2m^*\eta}\right)^{3/2}t^3 + \eta\left(\frac{\hbar^2}{2m^*\eta}\right)^2t^4 \implies 
     \frac{\Delta K(q)}{\eta q_0^4} =   t^2 - 2(\gamma/\gammacx)t^3 + t^4,
\end{align}
after introducing $\gammacx = 2\sqrt{\hbar^2\eta/2m^*}$.
We therefore see that $\Delta K(q) = K(q) - K(0)$ scales $\propto \eta q_c^4$. 

The minima of \eqref{eq:rescaled_kernel} are located at 
\begin{equation}
    2t - 6(\gamma/\gammacx)t^2 + 4t^3 = 0 \implies t = 0 \quad \text{and}\quad t_0 =\frac{q_0}{q_c} =  \frac{3}{4}\left(\frac{\gamma}{\gammacx} \pm \frac{\sqrt{\gamma^2 - \gammasp^2}}{\gammacx}\right),
\end{equation}
where $\gammasp = (4/3)\sqrt{\hbar^2\eta/m^*} = \sqrt{8}\gammacx/3$.
We have therefore found that the finite momentum solution exits only above the spinodal line $\gamma = \gammasp$. Along this line $t_0 = 1/\sqrt{2}$, while, for $\gamma =\gammacx$, the solution is $t_0 = 1$. Plugging this into Eq.\ \eqref{eq:rescaled_kernel}, one finds that $\Delta K(t_0 = 1) = 0$: we can therefore identify $\gammacx$ with the coexistence line where $\Delta K = 0$.

\subsection{Curvature and anisotropy of $K(q)$ around $q_i$}\label{app:scaling_analysis_1}

The kernel $K(\qb)$ can be expanded to quadratic order around $\qb_i$ with the help of the coefficients 
\begin{align}\label{eq:eigenvalues}
    C_\parallel = \left(\hbar^2/2m^*\right) -3q_0(\gamma - 2\eta q_0), \quad C_\perp &= \left(\hbar^2/2m^*\right) +q_0(3\gamma + 2\eta q_0),
\end{align} 
using that $q^3 \cos3\thetaq = q_x^3 - 3q_xq_y^2$. We remark that $C_\perp > C_\parallel$ always holds. 

For the sake of simplicity, it is convenient to work with the combination
\begin{equation}\label{eq:eigenvalues_bis}
    \Cbar = \frac{C_\perp +C_\parallel}{2} = \frac{\hbar^2}{2m^*} + 4\eta q_0^2,\quad \gamma = \frac{C_\perp - C_\parallel}{2\Cbar} = \frac{3\gamma q_0 -2\eta q_0^2}{\hbar^2/2m^* + 4\eta q_0^2},
\end{equation}
that describe the average curvature and anisotropy of $K(\qb)$ around $\qb_i$.

In terms of the rescaled variable introduced in the previous section, the coefficients of Eqs.\ \eqref{eq:eigenvalues} and \eqref{eq:eigenvalues_bis} read
\begin{equation}
 \frac{C_\parallel}{\eta q_c^2} = 1 - 6t_0\left(\frac{\gamma}{\gammacx} - t_0\right),\quad
 \frac{C_\perp}{\eta q_c^2} = 1 + 6t_0\left(\frac{\gamma}{\gammacx} + \frac{t_0}{3}\right),\quad \frac{\Cbar}{\eta q_c^2} =  1 + 4t_c^2,
\end{equation}
and 
\begin{equation}
  a = \frac{6(\gamma/\gammacx)t_0-2t_0^2}{1+4t_0^2}.
\end{equation}
We see that the coefficients $C_\parallel,C_\perp,\Cbar$ scale with $\eta q_c^2$, while $a$ depends only on $\gamma/\gammacx$ and changes from $a(\gammasp) = 1 $ to $a(\gammacx) =4/5$ and $ a(\gamma\gg \gammasp) \approx  1/2$ upon increasing the cubic term $\gamma/\gammacx$.

\section{Generality of the order parameter}\label{app:order_parameter}
Consider the combination of three condensates with distinct center-of-mass momenta $\qb_j = q_0\,\hat{\bm e}(\theta_j)$
 \begin{equation}
    \psi = \sum_{j=1}^3 \psi_j e^{i \qb_j\cdot \rb},
\end{equation}
with $|\psi_j|=|\psi_c|$.
Here, we show that the three phases $\varphi_j=\arg\psi_j$ do not play a physical role and can be absorbed into a redefinition of the global phase of the condensate and in a translation of the origin of the vortex lattice.
The combination of these transformations correspond to:
\begin{equation}
    \psi(\rb)\to \psi(\rb+\db)e^{i\chi} =|\psi_c|e^{i\chi} \left[e^{i\varphi_1}e^{i\qb_1\cdot (\rb+\db)}+e^{i\varphi_2}e^{i\qb_2\cdot (\rb+\db)}+e^{i\varphi_3}e^{i\qb_3\cdot (\rb+\db)}\right].
\end{equation}
By choosing the translation vector $\db$ as
\begin{equation}
    \db = -\left[\left(\frac{\varphi_1+\chi}{2\pi}\right)\ab_1+\left(\frac{\varphi_2+\chi}{2\pi}\right)\ab_2\right],
\end{equation}
with $\ab_j$ such that $\ab_j\cdot\qb_l=2\pi\delta_{jl}$ and $j,l=1,2$.
The order parameter then becomes
\begin{equation}
    \psi(\rb)\to \psi(\rb+\db)e^{i\chi} =|\psi_0|\left[e^{i\qb_1\cdot \rb}+e^{i\qb_2\cdot \rb}+e^{i\qb_3\cdot r+i\left(3\chi+\sum_{l=1}^{3}\varphi_l\right)}\right].
\end{equation}
Finally, by choosing the global phase $\chi$ as
\begin{equation}
    \chi=\frac{2\pi}{3}n-\frac{\varphi_1+\varphi_2+\varphi_3}{3},\quad\quad \text{with } n\in\mathbb Z,
\end{equation}
with $n\in\mathbb Z$ which concludes the proof. 
We remark that the global phase $\chi$ corresponds to the Goldstone mode associated with the phase of the superconducting order. In contrast, $\varphi_{2,3}$ represent phononic modes arising from the spontaneous breaking of translational symmetry.

 %We have shown that all the relative phases can be gauged away and do not induce any physical effect. Physically the global phase is associated to the phase Goldstone mode of the superconductor described by the compact variable $\chi\in[0,2\pi)$. On the other hand, the two component vector $\db$ is associated to phonons, which are invariant under a shift by $n\bb_2+m\bb_3$, i.e. compact variable defined in the unit cell. 

\section{Microscopic Modeling}\label{app:microscopic}

In this section, we discuss the microscopic model employed to describe tetralayer graphene. 

\subsubsection{Tight-binding model}

We use the lattice model for describing tetra-layer rhombohedral graphene~\cite{Jung2013,Zhang2010}:
\begin{equation}\label{sm:tetralayer}
    h =     \begin{pmatrix}
h^{(0)}_1 & h^{(1)} & h^{(2)} & 0  \\
h^{(1)\dagger} & h^{(0)}_2  & h^{(1)}& h^{(2)} \\
h^{(2)\dagger} & h^{(1)\dagger} & h^{(0)}_3 & h^{(1)} \\ 
0 & h^{(2)\dagger} &  h^{{(1)}\dagger} & h^{(0)}_4  
\end{pmatrix},
\end{equation}
where for simplicity we dropped the momentum dependency, $u_\ell =D (\ell-(N+1)/2)$ and: 
\begin{equation}
    h^{(0)}_l(\kb) = \begin{pmatrix}
        u_l & -t_0 f_{\kb} \\ 
        -t_0 f^*_{\kb} & u_l 
    \end{pmatrix},\quad
    h^{(1)}(\kb) = \begin{pmatrix}
        t_4 f_{\kb} & t_3 f^*_{\kb} \\ 
        t_1  & t_4 f_{\kb} 
    \end{pmatrix},\quad    h^{(2)} = \begin{pmatrix}
        0 & t_2/2  \\ 
        0 & 0 
    \end{pmatrix},\quad
    f_{\kb}=\sum^3_{j=1}e^{i\bm u_j\cdot\kb}.
\end{equation}
In the latter expression, we have introduced the vectors connecting the two sublattices $\bm \delta_j=a_G\mathcal R^{j-1}_{2\pi/3}(0,1)/\sqrt{3}$ and $a_G=0.246$nm.
Model parameters taken from~\cite{Dong_2024} are $t_0=3.1$eV, $t_1=380$meV and $(t_2,t_3,t_4)=(-21,290,141)$meV and $u_l$ intralayer potential reflecting the displacement field. The model features a characteristic length scale $\ell=\hbar v_F/t_1=a_G \sqrt{3}t_0/(2t_1)\approx7a_G\approx 2$nm.

\subsubsection{Interaction}

Our low energy theory is constructed employing valley and spin-polarized electrons with dispersion $\epsilon_{\kb}$ of the conduction band of tetra-layer graphene~\eqref{sm:tetralayer}:
\begin{equation}
    H_0=\sum_{\kb}\epsilon_{\kb}c^\dagger_{\kb}c_{\kb}.
\end{equation}
Furthermore, we employ a finite range attractive interaction: 
\begin{equation}
    H_{\rm int}=\frac{1}{2}\sum_{i\neq j}V(\rb_i-\rb_j),
\end{equation}
where $V(\rb)=-V_0e^{-r^2/(2a^2)}$ and in momentum space $V_{\qb}=-2\pi a^2 V_0 e^{-\qb^2a^2/2}$. In second quantization, we have: 
\begin{equation}
    H_{\rm int}=\frac{1}{2A}\sum_{\qb\kb\kb'}V_{\kb'-\kb}c^\dagger_{\kb+\frac \qb 2}c^\dagger_{\frac \qb 2-\kb}c_{\frac \qb 2-\kb'}c_{\kb'+\frac \qb 2},
\end{equation}
where we have introduced: 
\begin{equation}
    V_{\kb'-\kb} = -2\pi a^2 V_0 e^{-a^2(\kb-\kb')^2/2}.
\end{equation}
Employing the Jacobi-Anger expansion we find: 
\begin{equation}
    V_{\kb'-\kb}=-2\pi a^2 V_0 e^{-a^2k^2/2} e^{-a^2k'^2/2}\sum_{n=-\infty}^{\infty} I_n(kk'a^2) e^{in\theta}e^{-in\theta'},
\end{equation}
where $I_n(x)$ is the modified first Bessel function. 
In the limit of short-ranged attraction $|kk'a^2|\ll 1$ and we find:
\begin{equation}
    I_n(kk'a^2)\approx \frac{(kk'a^2)^n}{2^n n!}.
\end{equation}
The latter expansion is justified when the interaction range $a$ is smaller than the characteristic interparticle separation $d=1/\sqrt{n_{e}}\approx10$\,nm determining the relevant momentum scale. 
Keeping only the leading contribution to the momentum expansion:\begin{equation}\label{pairing_pseudopotential}
    V_{\kb'-\kb}\approx-\pi a^2 V_0\sum_{n=\pm} (\kb_n a) ({\kb'}_n^*a) = -\pi a_G^2(a/a_G)^4 V_0\sum_{n=\pm} (\kb_na_G) ({\kb'}_n^*a_G).
\end{equation}

Without losing generality, in the following we perform calculations keeping the general structure $V_{\kb,\kb'}=-g\Delta_{\kb}\Delta_{\kb'}^*$ with $\Delta_{\kb}$ gap function.
We then perform specific toy model calculations under the assumption that the normal state favors $p$-wave symmetry with a well-defined chirality, characterized by$L_z=\pm1$ depending on the valley $K/K'$~\cite{Geier_2024}. The interaction takes the form $V_{\kb'-\kb}=-g\kb_\pm{\kb}'_{\mp}$ with $g=\pi (a/a_G)^4a_G^2 V_0$.

\subsection{Derivation of the Ginzburg-Landau theory}\label{app:GL_microscopic}
\begin{figure}
    \centering
\includegraphics[width=0.45\linewidth]{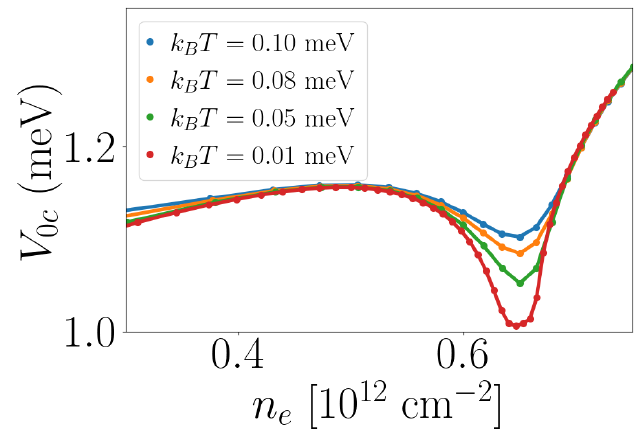}
    \caption{Critical coupling strength $V_{0c}$~\eqref{critical_coupling} as a function of the temperature for different carrier concentrations and $D=30$meV. }
    \label{fig:critical_coupling}
\end{figure}

The action describing the effective theory $H=H_0+H_{\rm int}$ reads: 
\begin{equation}\label{sm:action}
    S = \int^\beta_0d\tau \left[\sum_{\kb} c^\dagger_{\kb}(\partial_\tau-\xi_{\kb})c_{\kb} +\frac{1}{2}\sum_{\qb\kb}\psi(\qb) \Delta_{\kb} c^\dagger_{\kb+\qb/2} c^\dagger_{\qb/2-\kb} + h.c. + A \sum_{\qb}\frac{|\psi(\qb)|^2}{2g} \right],
\end{equation}
where $\xi=\epsilon-\mu$, $\psi(\qb)$ is the Hubbard-Stratonovich field associated to the pairing field and $\Delta_{\kb}$ the gap function $\Delta_{-\kb}=-\Delta_{\kb}$. We integrate the fermions and expand the theory to second order in the pairing field $\psi(\qb)$: 
\begin{equation}\label{quadratic_free_energy}
     F^{(2)}_\psi = A \sum_{\qb}\frac{|\psi(\qb)|^2}{2g}-\frac{T}{2}\sum_{\qb}|\psi(\qb)|^2\sum_{\omega_n}\int_{\kb} |\Delta_\kb|^2 G(\kb+\qb/2,\omega_n)G(\qb/2-\kb,-\omega_n)%+\mathcal O(|\psi|^4)
     =\sum_{\qb}\psi(\qb)^*K(\qb)\psi(\qb),
\end{equation}
where $\omega_n=i\pi(2n+1)k_BT$, $G^{-1}(\kb,\omega_n)=(i\omega_n-\xi_{\kb})$ and we have introduced $K(\qb)$ shown in Fig.~\ref{fig:box_diagrams}a) and given by: 
\begin{align}
    K(\qb)&=\frac{1}{2g}-\frac{T}{2}\sum_{\omega_n}\int_{\kb} |\Delta_\kb|^2 G\left(\kb+\frac{\qb}{2},\omega_n\right)G\left(\frac{\qb}{2}-\kb,-\omega_n\right)\nonumber\\
    &=\frac{1}{2g}-\frac{\Omega}{4}\int \frac{d^2\kb}{(2\pi)^2}\frac{|\Delta_\kb|^2}{\xi_{\qb/2+\kb}+\xi_{\qb/2-\kb}}\left(\tanh\frac{\xi_{\qb/2+\kb}}{2k_B T}+\tanh\frac{\xi_{\qb/2-\kb}}{2k_B T}\right),
\end{align}
where $\Omega=|\bm a_1\times \bm a_2|=\sqrt{3} a^2_G/2$ is the area of the unit cell. 
By symmetry arguments, for a small center-of-mass momenta we can expand $K(\qb)$ to find Eq.~\eqref{eq:propagator} in the main text.
%The latter expansion accurately describes the numerical results, as shown by the fit in Fig.~\ref{fig3}.  
\begin{figure}
    \centering
    \includegraphics[width=0.7\linewidth]{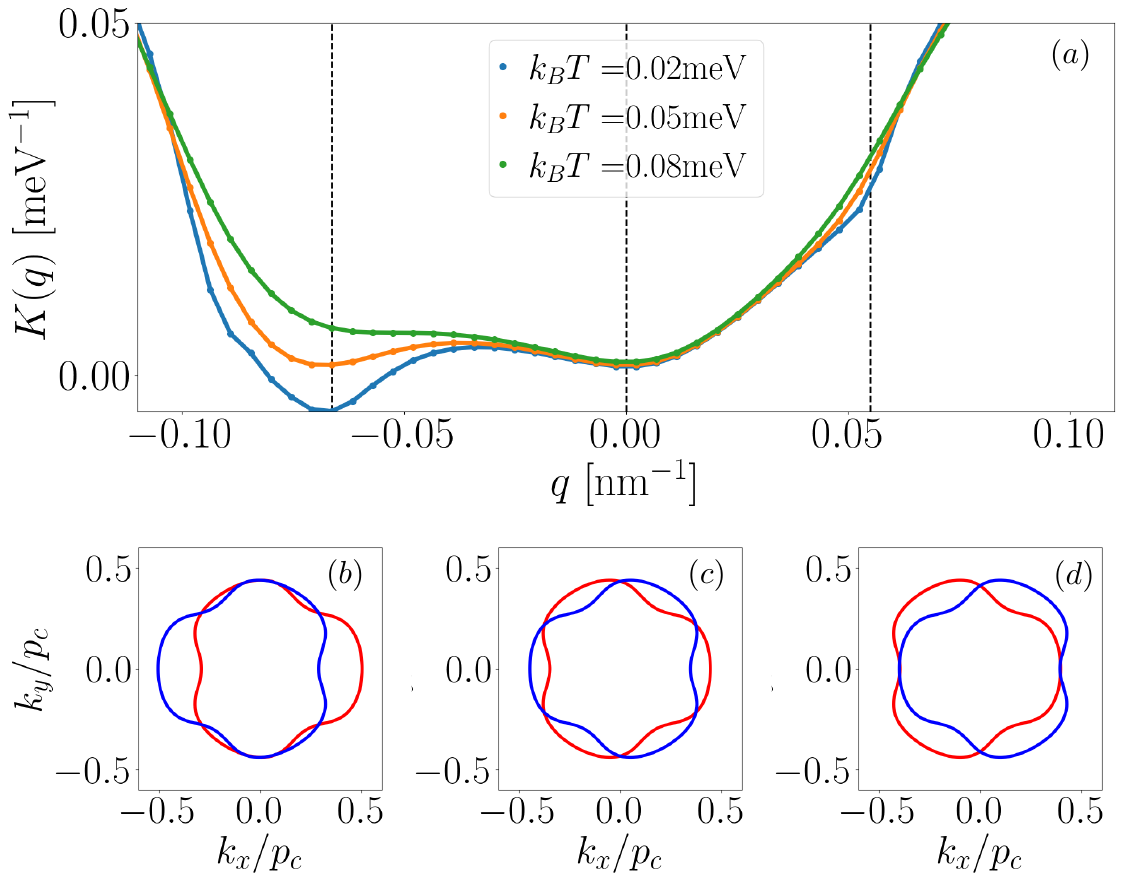}
    \caption{Panel $(a)$ shows $K(\qb)$ along $q_x$ for different temperatures at filling factor $n=0.45\times 10^{12}$cm$^{-2}$ and $D=30$meV. We employed $V_0=1.15$meV and $a=20a_G$. Panels $(b)$, $(c)$ and $(d)$ show the electrons (red) and hole (blue) Fermi surface with net center-of-mass momenta $\qb$ correspondent to the three dashed vertical lines in panel $(a)$. Momenta are measured in unit of $p_c=t_1/\hbar v_F$. }
    \label{fig:quadratic_coefficient_sm}
\end{figure}

We emphasize that the absence of the Cooper log singularity requires that the coupling constant $g$ is larger than a critical threshold in order for the superconductivity to take place. We provide an estimate of the critical coupling strength $g_c$ for $\qb=0$ superconductivity, defined as $K(\qb=0)=0$:
\begin{equation}
    g_c(T,n_e) =\left[\frac{\Omega}{2}\int \frac{d^2\kb}{(2\pi)^2}\frac{|\Delta_\kb|^2}{\xi_{\kb}+\xi_{-\kb}}\left(\tanh\frac{\xi_{\kb}}{2k_B T}+\tanh\frac{\xi_{-\kb}}{2k_B T}\right)\right]^{-1}, 
\end{equation}
where the numerical analysis has been performed utilizing the form factor $\Delta_\kb=k_x+ik_y$ characteristic of a chiral $p$-wave superconductor~\cite{Read_2000}. 
Exploiting the relation in Eq.~\ref{pairing_pseudopotential}, we find that the critical interaction strength reads: 
\begin{equation}\label{critical_coupling}
    V_{0c}(T,n_e)=\frac{a^2_Gg(T,n_e)}{\pi a^4}.
\end{equation}
Fig.~\ref{fig:critical_coupling} displays $V_{0c}$ as a function of the filling factor for different values of the temperature where we set $a=20 a_G\approx 5$nm which remains smaller than the interparticle separation $d\ge 10$nm.

Fig.~\ref{fig:quadratic_coefficient_sm}$(a)$ shows the evolution of $K(\qb)$ for different values of the temperature at filling factor $n_e=0.45$. Figs.~\ref{fig:quadratic_coefficient_sm}$(b)$, $(c)$ and $(d)$ depict the electron $\xi_{\qb/2-\kb}$ and hole $-\xi_{\qb/2-\kb}$ Fermi surfaces for three different center-of-mass shifts $\qb$. The resonant condition is achieved for Fig.~\ref{fig:quadratic_coefficient_sm}$(a)$ where the Cooper pair has net center-of-mass momentum $\qb_1$. We remark that this condition is also achieved for the $C_{3z}$ symmetric momenta $\qb_{2}=C_{3z}\qb_1$ and $\qb_{3}=C^2_{3z}\qb_1$.

\begin{figure}
    \centering
    \includegraphics[width=.8\linewidth]{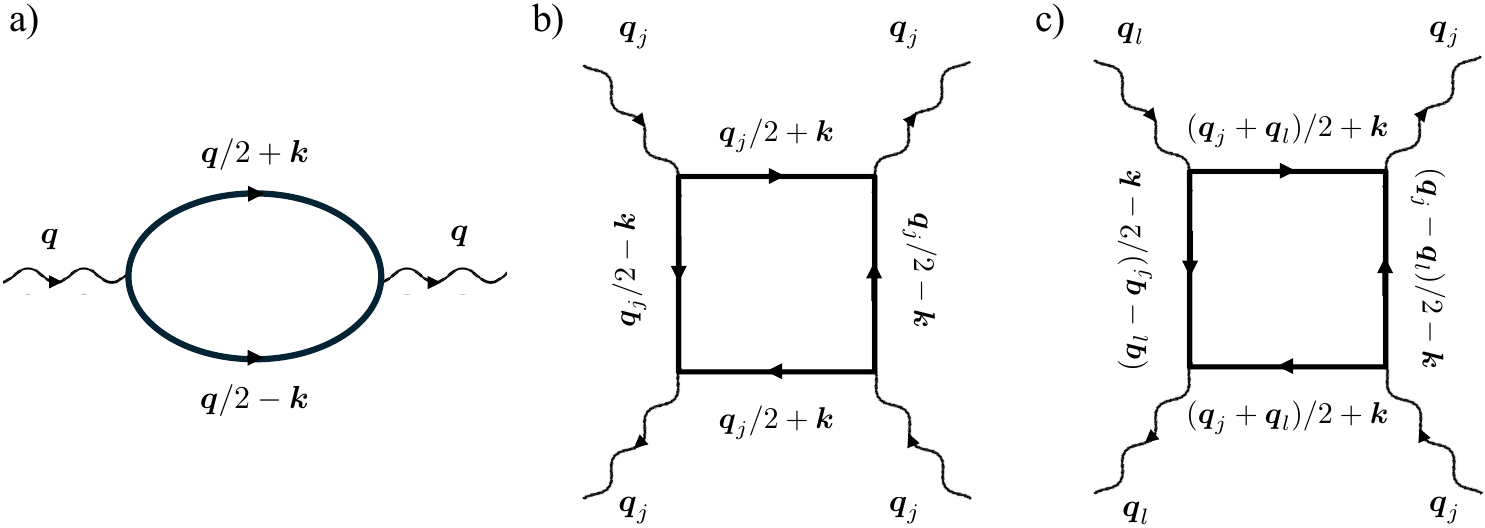}
    \caption{Panel a) shows the bubble diagram for the quadratic part $K(\qb)$ of the superconducting action. Panels b) and c) show the box diagrams for the $|\psi|^4$ contribution to the free energy, with $j=l$ in b) and $j\neq l$ in c). Incoming and outgoing wavy lines are $\psi^*(\qb_{j/l})$ and $\psi(\qb_{j/l})$, respectively. Each vertex $\Delta_{\kb}$ carries momentum given by the difference between the momenta of the incoming and outgoing fermionic lines.}
    \label{fig:box_diagrams}
\end{figure}

To differentiate between different finite momentum condensates, we calculate the quartic contribution $\propto|\psi|^4$ to the action of the paring field. This is expressed in terms of box diagrams that includes 4 vertices and 4 internal legs. Each vertex carries a momentum label $\qb_{j}$ with $j=1,2,3$. Particle number conservation implies that out of 4 vertices two are $\psi^*(\qb_{j_{1/3}})$ and the other two $\psi(\qb_{j_{2/4}})$. Finally, momentum conservation enforces $\qb_{j_1}+\qb_{j_3}=\qb_{j_2}+\qb_{j_4}$. To start with, we consider the diagonal contribution where $\qb_{j_1}=\qb_{j_3}$ and, therefore, $\qb_{j_2}=\qb_{j_4}$. The resulting contribution is shown in Fig.~\ref{fig:box_diagrams}b) and reads: 
\begin{equation}\label{sm:beta}
    \beta=\frac{T}{4}\sum_{\omega_n}\int_{\kb}|\Delta_\kb|^4 G\left(\frac{\qb_j}{2}+\kb,\omega_n\right)^2G\left(\frac{\qb_j}{2}-\kb,-\omega_n\right)^2,
\end{equation}
where as a result of the $C_{3z}$ symmetry the integral does not depend on $j=1,2,3$.
The off-diagonal term shown in Fig.~\ref{fig:box_diagrams}c) takes the form: 
\begin{equation}\label{sm:betaprime}
    \beta'=\frac{T}{2}\sum_{\omega_n}\int_{\kb}\left|\Delta_{\kb+\qb_j/2}\right|^2\left|\Delta_{\kb+\qb_l/2}\right|^2 G\left(\frac{\qb_j+\qb_l}{2}+\kb,\omega_n\right)^2G\left(\frac{\qb_j-\qb_l}{2}-\kb,-\omega_n\right)G\left(\frac{\qb_l-\qb_j}{2}-\kb,-\omega_n\right),
\end{equation}
where the coefficient does not depend on the pair $j\neq l$ due to $C_{3z}$ and $TM_x$ symmetries. 

Summing together the fourth order contribution with $F^{(2)}_\psi$~\eqref{quadratic_free_energy}, we find the energy of the finite momentum condensate:
\begin{equation}
    F_\psi = \alpha\sum_{j=1}^{3} |\psi_j|^2+\beta\sum_{j=1}^{3}|\psi_j|^4+\beta'\sum_{i\neq j}|\psi_i|^2|\psi_j|^2.
\end{equation} 

\begin{figure}
    \centering
    \includegraphics[width=.7\linewidth]{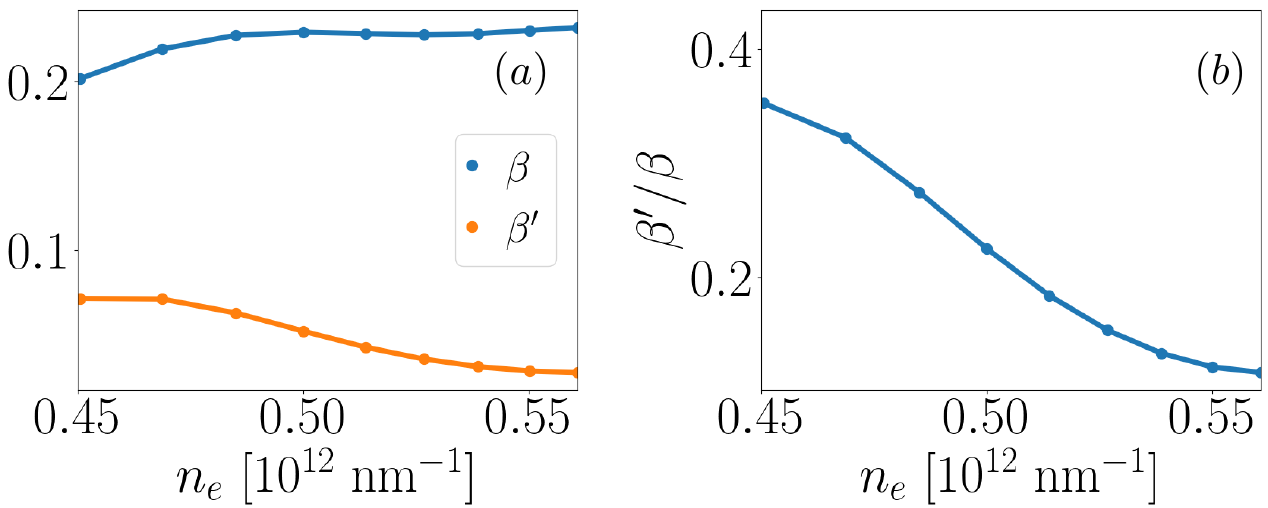}
    \caption{Panel $(a)$ shows the quartic coefficients $\beta$ and $\beta'$ as a function of the density $n_e$ expressed in arbitrary units. Panel $(b)$ shows the ratio $\beta'/\beta$. In this calculation, we employ $k_BT=.02$meV, $D=30$meV and the momentum $q_0$ is set by the absolute minimum of $K(\qb)$. }
    \label{fig:beta_beta_prime}
\end{figure}

Before moving on, we perform the integrals over Matsubara frequencies in Eqs.~\eqref{sm:beta} and~\eqref{sm:betaprime} and we provide numerical results for $\beta$ and $\beta'$.
The first integral reads:
\begin{equation}
    T\sum_{\omega_n}G\left(\frac{\qb_j}{2}+\kb,\omega_n\right)^2G\left(\frac{\qb_j}{2}-\kb,-\omega_n\right)^2=\frac{\partial_\epsilon f|_{\xi_{\qb_j/2+\kb}}+\partial_\epsilon f|_{-\xi_{\qb_j/2-\kb}}}{(\xi_{\qb/2+\kb}+\xi_{\qb/2-\kb})^2}+\frac{\tanh\frac{\xi_{\qb_j/2+\kb}}{2k_B T}+\tanh\frac{\xi_{\qb_j/2-\kb}}{2k_B T}}{(\xi_{\qb/2+\kb}+\xi_{\qb/2-\kb})^3}.
\end{equation}
The second contribution takes the form: 
\begin{equation}
    \begin{split}
        &T\sum_{\omega_n} G\left(\kb_1,\omega_n\right)^2G\left(\kb_2,-\omega_n\right)G\left(\kb_3,-\omega_n\right)=\frac{\partial_{\epsilon}f|_{\xi_{\kb_1}}}{(\xi_{\kb_1}+\xi_{\kb_2})(\xi_{\kb_1}+\xi_{\kb_3})}+\frac{f(-\xi_{\kb_2})}{(\xi_{\kb_2}+\xi_{\kb_1})^2(\xi_{\kb_3}-\xi_{\kb_2})}\\
        &+\frac{f(-\xi_{\kb_3})}{(\xi_{\kb_3}+\xi_{\kb_1})^2(\xi_{\kb_2}-\xi_{\kb_3})} - \frac{(2\xi_{\kb_1}+\xi_{\kb_2}+\xi_{\kb_3})f(\xi_{\kb_1})}{(\xi_{\kb_1}+\xi_{\kb_2})^2(\xi_{\kb_1}+\xi_{\kb_3})^2},
    \end{split} 
\end{equation}
where $f(E)=1/(e^{E/k_BT}+1)$ and we introduced $\kb_1=\kb+(\qb_j+\qb_l)/2$ and $\kb_{2/3}=\pm(\qb_j-\qb_l)/2-\kb$. 
Figure~\ref{fig:beta_beta_prime} presents the numerical evaluation of $\beta$ and $\beta'$ in the relevant regime, where a pairing instability with net center-of-mass momentum occurs. 
In this regime, we found $\beta>\beta'>0$ leading to the formation of the VAL. 

% We summarize some useful integrals:
% \begin{equation}\begin{split}
%     J_d(\kb)=T\sum_{i\epsilon}\frac{1}{(i\epsilon-\xi_{\kb+\qb/2})^2(i\epsilon+\xi_{\qb/2-\kb})^2}=&\frac{f'(\xi_{\kb+\qb/2})+f'(-\xi_{\qb/2-\kb})}{(\xi_{\kb+\qb/2}+\xi_{\qb/2-\kb})^2}\\
%     &+\frac{1}{(\xi_{\kb+\qb/2}+\xi_{\qb/2-\kb})^3}\left[\tanh\frac{\xi_{\kb+\qb/2}}{2k_BT}+\tanh\frac{\xi_{\qb/2-\kb}}{2k_BT}\right].
% \end{split}\end{equation}
% In addition, we also have: 
% \begin{equation}\begin{split}
%     J_o(\kb)=&T\sum_{i\epsilon}\frac{1}{(i\epsilon-\xi_{\kb+\frac{\qb+\qb'}{2}})^2(i\epsilon+\xi_{\frac{\qb-\qb'}{2}-\kb})(i\epsilon+\xi_{\frac{\qb'-\qb}{2}-\kb})}=\frac{f'(\xi_{\kb+\frac{\qb+\qb'}{2}})}{(\xi_{\kb+\frac{\qb+\qb'}{2}}+\xi_{\frac{\qb-\qb'}{2}-\kb})(\xi_{\kb+\frac{\qb+\qb'}{2}}+\xi_{\frac{\qb'-\qb}{2}-\kb})}\\
%     &+\frac{1}{(\xi_{\kb+\frac{\qb+\qb'}{2}}+\xi_{\frac{\qb-\qb'}{2}-\kb})^2}\left[\frac{f(-\xi_{\frac{\qb-\qb'}{2}-\kb})}{\xi_{\frac{\qb'-\qb}{2}-\kb}-\xi_{\frac{\qb-\qb'}{2}-\kb}}-\frac{f(\xi_{\kb+\frac{\qb+\qb'}{2}})}{\xi_{\kb+\frac{\qb+\qb'}{2}}+\xi_{\frac{\qb'-\qb}{2}-\kb}}\right]\\
%     &+\frac{1}{(\xi_{\kb+\frac{\qb+\qb'}{2}}+\xi_{\frac{\qb'-\qb}{2}-\kb})^2}\left[\frac{f(-\xi_{\frac{\qb'-\qb}{2}-\kb})}{\xi_{\frac{\qb-\qb'}{2}-\kb}-\xi_{\frac{\qb'-\qb}{2}-\kb}}-\frac{f(\xi_{\kb+\frac{\qb+\qb'}{2}})}{\xi_{\kb+\frac{\qb+\qb'}{2}}+\xi_{\frac{\qb-\qb'}{2}-\kb}}\right],
% \end{split}\end{equation}
% where $f(E)=1/(e^{E/k_B T}+1)$.

\subsection{Continuum BdG Hamiltonian of the VAL phase}\label{sec:BdG}

In this section we derive the microscopic BdG Hamiltonian describing the VAL phase. 
Starting from the superconducting order parameter $\psi(\rb)=|\psi_c|\sum_{j=1}^{3}e^{i\qb_j\cdot\rb}$, the continuum BdG Hamiltonian~\cite{Franz_2000,Vafek_2001} reads:  
\begin{equation}
    H_{\rm BdG} = \int d^2\rb \left[\Psi^\dagger\xi_{-i\nabla}\Psi+\frac{1}{4}\Psi^\dagger\{\psi(\rb),-i\bar \partial\}\Psi^\dagger +h.c.\right],
\end{equation}
where $\bar\partial = \partial_x+i\partial_y$ ($-i\bar\partial e^{i\qb\cdot\rb}=e^{i\qb\cdot\rb}(\qb_+-i\bar\partial)$), $\{\cdot,\cdot\}$ is the anticommutator. 

We now introduce the reciprocal lattice vectors $\gb=n_1\gb_1+n_2\gb_2$ where $\gb_1=\qb_1-\qb_2$ and $\gb_2=\qb_2-\qb_3$ ($\gb_3=\qb_3-\qb_1$). Given $\kb\in\Pi$ in the first Brillouin zone $\Pi$, the BdG Hamiltonian expanded in plane waves reads:
\begin{equation}\begin{split}
    H_{\rm BdG}(\kb)&=\frac{1}{2}\sum_{\gb} \left[\xi_{\kb+\gb}c^\dagger_{\kb+\gb}c_{\kb+\gb}-\xi_{-\kb-\gb+\qb_1}c_{-\kb-\gb+\qb_1}c^\dagger_{-\kb-\gb+\qb_1}\right]\\
    &+\frac{1}{2}\sum_{j=1}^{3}\psi_j\sum_{\gb}\left(\kb+\gb-\frac{\qb_j}{2}\right)_+c^\dagger_{\kb+\gb}c^\dagger_{\qb_1-\bm Q_j-\gb-\kb}+ h.c.,
\end{split}\end{equation}
where $\bm Q_j=\qb_1-\qb_j$ and $\kb_\pm=k_x\pm ik_y$. 
Using the Nambu basis: 
\begin{equation}
    \Phi_{\gb}(\kb)=\begin{pmatrix}
        c_{\kb+\gb} \\
        c^\dagger_{\qb_1-\kb-\gb}
    \end{pmatrix},
\end{equation}
the Hamiltonian takes the compact form: 
\begin{equation}
    H_{\rm BdG}(\kb)=\frac 1 2\sum_{\gb}\Phi^\dagger_{\gb}(\kb)  h_{\gb\gb'}(\kb)\Phi_{\gb'}(\kb).
\end{equation}
In the latter expression we have:
\begin{equation}
    h_{\gb\gb'}(\kb)= \delta_{\gb\gb'}\begin{pmatrix}
        \xi_{\kb+\gb} & 0 \\ 
        0 & -\xi_{\qb_1-\kb-\gb}
    \end{pmatrix}+|\psi_c|\sum_{j=1}^{3}\delta_{\gb',\gb+\bm Q_j}\tau^+\left(\kb+\gb-\frac{\bm q_j}{2}\right)_++h.c.,
\end{equation}
where $\tau^\pm$ are raising and lowering matrices in the Nambu space.

\end{document}